\documentclass[multphys,vecphys]{svmult}

\usepackage{makeidx}     
\usepackage{graphicx}    

\begin{document}

\newcommand{\Btot}{B_\mathrm{tot}}
\newcommand{\Bobs}{B_{||}}
\newcommand{\Bpar}{B_{||}}
\newcommand{\Bperp}{B_{\perp}}
\newcommand{\Nobs}{N_\mathrm{obs}}
\newcommand{\Nperp}{N_{\perp}}
\newcommand{\Va}{V_\mathrm{NT}}
\newcommand{\Vntobs}{V_\mathrm{NTobs}}
\newcommand{\HI}{{\rm H\,\scriptstyle I}}
\newcommand{\HII}{{\rm H\,\scriptstyle II}}
\newcommand{\cmcube}{\,{\rm cm^{-3}}}
\newcommand{\cmsix}{\,{\rm cm^{-6}}}
\newcommand{\radm}{\,{\rm rad\, m^{-2}}}
\newcommand{\kms}{\,{\rm km\, s^{-1}}}
\newcommand{\pc}{\,{\rm pc}}
\newcommand{\kpc}{\,{\rm kpc}}
\newcommand{\EM}{{\rm EM}}
\newcommand{\RM}{{\rm RM}}
\newcommand{\DM}{{\rm DM}}
\def\degr{\hbox{$^\circ$}}
\def\arcmin{\hbox{$^\prime$}}
\def\arcsec{\hbox{$^{\prime\prime}$}}
\def\fdg{\hbox{$.\!\!^\circ$}}
\def\farcm{\hbox{$.\mkern-4mu^\prime$}}
\def\farcs{\hbox{$.\!\!^{\prime\prime}$}}

\def\simlt{\lower.5ex\hbox{$\; \buildrel < \over \sim \;$}}
\def\simgt{\lower.5ex\hbox{$\; \buildrel > \over \sim \;$}}

\title*{Magnetic Fields in Diffuse $\HI$ and Molecular Clouds}

\author{Carl Heiles\inst{1}\and
Richard Crutcher\inst{2}}
\authorrunning{C. Heiles and R. Crutcher}

\institute{Astronomy Department, University of California,
Berkeley
\texttt{cheiles@astron.berkeley.edu}
\and Astronomy Department, University of Illinois\newline
 \texttt{crutcher@uiuc.edu}}

\maketitle



\section{Introduction}
\label{intro}

The diffuse interstellar $\HI$ is the matrix within which many
molecular clouds reside and the medium that soaks up the energy injected
by sources such as supernovae and stellar winds.  This energy stimulates
turbulence in the $\HI$, which cascades up the turbulent wavenumber
spectrum.  The spectral wavelengths extend all the way down to scales
most easily quoted in Astronomical Units. $\HI$ and molecular clouds
enjoy a synergistic relationship, with turbulent energy, angular
momentum, magnetic fields, and matter flowing across the boundaries in
both directions. The molecular clouds form stars, which in turn act as
energy sources to round the circle and make star formation a feedback
process.

\noindent Fortunately for us who study magnetic fields, the neutral medium
isn't really neutral and, as a consequence, flux freezing applies.  In
diffuse $\HI$ the minimum free electron fraction is, at minimum, equal
to that of heavy elements that have ionization potential less than that
of $\HI$ ($\simgt 10^{-4}$) because even in the dark reaches of space
there are plenty of starlight photons available to keep any such element
ionized. As a crude approximation we can model a piece of the
interstellar gas as a giant inductor, for which the timescale $\tau$ for
decay of a current (and its associated magnetic field) is the inductance
divided by the resistance; this, in turn, goes as $\tau \propto {L^2 /
\eta}$, where $L$ is the length scale and $\eta$ the resistivity.
Even with the low fractional ionization, $L$ dominates and timescales
for decay are always long in diffuse $\HI$. In dense molecular clouds
starlight is excluded and the free electrons come from cosmic-ray
ionization of H; the fractional ionization is small enough that slow
leakage of frozen magnetic flux allows the clouds to gradually evolve.

With flux freezing, the magnetic field becomes one of the four most
important forces on the diffuse gas. The others are gas pressure,
cosmic-ray pressure, and gravity.  Gravity dominates on the
largest scales, e.g.\ by keeping the gas pulled down as part of the
Galactic plane; it also dominates during star formation, of course.  On
all other scales the gas responds only to the three pressure forces. The
gas and cosmic rays are connected by the field, so they form a coupled
system. The field is a -- perhaps {\em the} -- major player.

One determines the field strength in the diffuse interstellar gas in several ways.
Each method has its own idiosyncrasies and provides values that are
biased either up or down. Beck et al. (2003)
is {\em required reading} to understand these biases.  Synchrotron
emissivity provides a volume average of $\langle B^x \rangle ^{1/x}$,
where $1.9 \simlt x \simlt 3.9$ depending on whether one assumes the
electron cosmic-ray spectrum or energy equipartition (Beck 2001).
Comparing pulsar rotation and dispersion measures provides a field
strength in the diffuse Warm Ionized Medium (WIM). Zeeman splitting
provides the field strength in the $\HI$.

Combining these estimates gives a typical magnetic field strength $\sim
6 \pm 2\,\umu$G (Beck 2001),
which is equivalent to a gas pressure $\tilde P \equiv {P/k} \sim
10400\cmcube$\,K.  This is about three times the typical ISM thermal
gas pressure of $\sim 3000\cmcube$\,K (Jenkins \& Tripp 2001, Wolfire et
al. 2003),
and is comparable to the other important interstellar energy densities,
namely turbulence and cosmic rays. These pressures must add to provide
hydrostatic support for the gas layer, estimated to be $P_\mathrm{tot}
\approx 28000\cmcube$\,K at $z = 0$ (Boulares \& Cox 1990).
Clearly, thermal pressure is a minority player; turbulence, cosmic
rays, and the magnetic field dominate. One cannot hope to understand
the interstellar medium without understanding the role of the magnetic
field.  Moreover, the crucial star formation feedback process is
regulated, or stimulated, or at least greatly affected, by the magnetic
field.

Magnetism makes its effects very clear in supernova shocks. These
shocks compress both the gas and the field. As the gas cools behind the
shock, it does so at roughly constant pressure, so its density
increases. Concomitantly, the field strength increases because of flux
freezing. Magnetic pressure increases as $B^2$, so eventually the
magnetic pressure prevents the gas from condensing further. This
limits the compression of gas behind the shock and over the latter
stages of its evolution the magnetic field greatly increases the shell
thickness relative to the idealized nonmagnetic case. Moreover, on the
full scale of the shell the magnetic field acts as a retarding force,
increasing the deceleration of the shell and reducing its final size
(Tomisaka 1990, Ferri\`ere et al. 1991, Slavin \& Cox 1992).
Also, the strong field can inhibit the production of worms (Heiles 1984)
and chimneys (Norman \& Ikeuchi 1989).

For the study and interpretation of magnetic fields, the size scale is
paramount. At the largest scales within galaxies, the global scale, the
issue is field generation and maintenance, and the underlying questions
are ``Primordial field or dynamo?'' and ``What kind of dynamo?''. These
questions are addressed by size scales ranging down to spiral arms.  At
smaller sizes we have the field in individual interstellar diffuse
structures, which are shaped by point energy injection and condensation
onto molecular clouds.  At yet smaller scales we have molecular clouds,
especially those that contain protostellar cores. At the smallest
scales we have regions where stars have formed.

This review concentrates on the magnetic field at intermediate and small
size scales, i.e.\ diffuse $\HI$ structures and molecular clouds and cores.
See Beck (2001) for discussion of magnetic fields on larger scales.

Our chosen size range is where energy input to the ISM occurs and where
energy is transferred by turbulence to smaller scales and across cloud
boundaries.  There are three, and only three, established\footnote{Use of
the difference in line widths between neutral and ionized species to infer
the angle between the line of sight and the magnetic field (Houde et al. 2002)
and Faraday screens in dark-cloud envelopes (Wolleben \& Reich 2004) are
possible additional techniques that have not yet been fully accepted.} tracers
for the field at these scales: polarization from aligned dust grains, which
both absorb starlight and emit in the far-infrared, linear polarization
of spectral lines, and Zeeman splitting of
spectral lines.  We will briefly include starlight polarization in
Sect.~\ref{starlightpol}, concentrate on Zeeman splitting of the 21-cm
line in Sects.~\ref{binabs} and \ref{bemission}, and discuss magnetic fields
in molecular clouds starting with Sect.~\ref{molecularclouds}.

One major focus of this review is the magnetic field in the diffuse $\HI$.
The $\HI$ resides in two thermal phases, the Cold Neutral Medium (CNM)
and the Warm Neutral Medium (WNM), each containing roughly half of the
total $\HI$. Classically, we imagine these as points of stable
isobaric thermodynamic equilibrium (Field 1965, McKee \& Ostriker 1977),
with the temperatures differing by about two orders of magnitude.  The
CNM does, in fact, reside in the classical stable thermal equilibrium.
However, the WNM is buffeted by many agents on a range of timescales, so
much so that at least 50\% of the WNM has temperature smaller than
5000\,K, meaning that it is {\em not} thermally stable (Heiles \&
Troland 2003).
The WNM, being of much higher temperature and lower density, occupies
the lion's share of the interstellar volume, roughly half the volume in
the Solar vicinity (Heiles 2000b).
$\HI$ Zeeman splitting measurements refer almost exclusively to the
CNM: the line widths of the WNM are large, and when combined with $\HI$
angular structure the instrumental effects have so far prohibited
reliable measurements.

The other major focus is the magnetic field in molecular clouds. The
most important goal is to understand the role that magnetic fields play
in the fundamental astrophysical process of star formation. One view
is that self-gravitating clouds are supported against
collapse by magnetic fields, with ambipolar diffusion reducing
support in cores and hence driving star formation
(Mouschovias \& Ciolek 1999). The other view is that clouds form and
disperse by the operation of compressible turbulence (e.g., Elmegreen 2000),
with clumps sometimes becoming gravitationally bound and collapsing to
form stars. The issue of which (if either) of these paradigms for the
evolution of molecular clouds and the formation of stars is correct
is currently unresolved. We describe the state of observations of
magnetic fields in molecular clouds and how these data may be used
to test predictions of the two star formation paradigms.

\section{Measuring the Magnetic Field in Diffuse $\HI$ and Molecular Clouds}
\label{diffusefield}

\subsection{Polarization of Starlight by Magnetically Aligned Grains}
\label{starlightpol}

Polarization of starlight holds the enviable position of being the
means by which the interstellar magnetic field was discovered (see
Davis \& Greenstein (1951) for references and the original theory of
grain alignment). Their alignment mechanism involves charged, spinning
interstellar grains whose angular momentum vector component parallel to
the field is damped by paramagnetic relaxation. The theory evolved with
the introduction of superthermal spins and internal damping from
Barnett relaxation (Purcell 1979, Purcell \& Spitzer 1971).
The theory continues to evolve as more exotic effects are uncovered
(see Lazarian (2003) for a comprehensive review devoted exclusively to
grain alignment; also see Draine (2003) and references quoted therein).
In principle, the starlight polarization can be either parallel or
perpendicular to $\Bperp$, the field on the plane of the sky. However,
empirically the polarization is parallel to the field, as revealed by
polarization in diffuse regions near the Galactic plane: $\Bperp$ is
parallel to the plane as expected for the Galactic-wide field.

\begin{figure}[!p]
\centering
\includegraphics[bb = 172 124 432 664,angle=-180,width=3.5in]{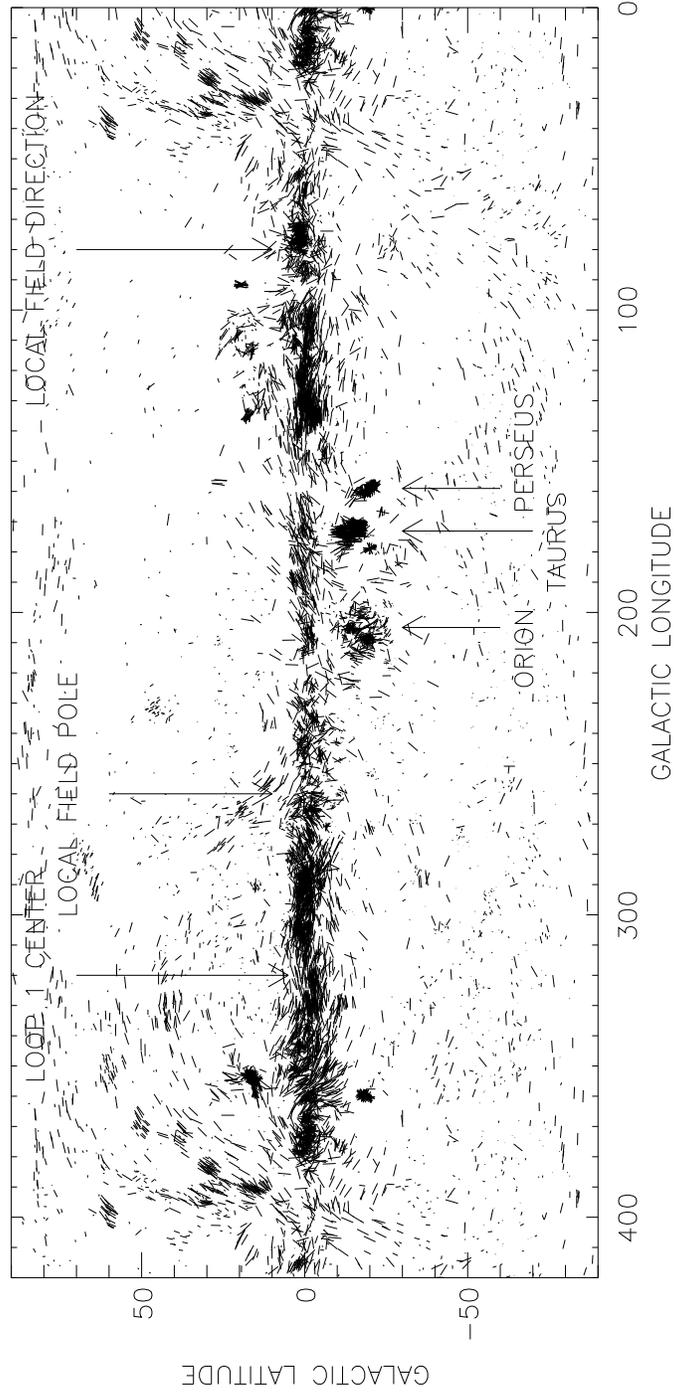}
\caption{Starlight polarization of 8662 stars. The orientation of each
star's polarization is indicated by a short line whose length $L$ in
great-circle degrees is $L = [4 < 2P]\degr$, where $P$ is the
percentage polarization; for $L$, we plot whichever of the two
quantities is smaller. \label{lbpol} }
\end{figure}

Starlight polarization is produced by aligned dust that selectively
absorbs one direction of linear polarization more than the orthogonal
one. This makes the fractional polarization proportional to the
extinction -- we can't have polarization without extinction! Commonly,
maps represent starlight polarization with lines whose direction is
that of the polarization and whose length is proportional to the
fractional polarization. The eye notices the long lines, which
emphasize high extinction; these stars tend to be more distant. This
is normally not the kind of bias one wants. For example, if we are
interested in the nearby field structure, it is better to make all
lines the same length. Accordingly, in our Fig.~\ref{lbpol}, we
de-emphasize distant or high-extinction stars by placing an upper
limit on the length of the lines.

The fractional starlight polarization also increases as the field
becomes perpendicular to the line of sight. The dependence is
$( \Bperp / \Btot )^2$. From our discussion in
Sect.~\ref{univarb}, for randomly oriented fields this ratio has mean
value 0.67 and median 0.87. With these high numbers, most of the
regions have a high ratio, so in a statistical sample the fractional
polarization is relatively weakly affected by the tilt of the magnetic
field. Statistically, extinction is much more important in determining
the fractional polarization.

Figure~\ref{lbpol} shows the polarization of 8662 stars from the
compilation of known catalogs (Heiles 2000a). The orientation of
each star's polarization is indicated by a short line whose length $L$
in great-circle degrees is $L = [4 < 2P]\degr$, where $P$ is the
percentage polarization; we cap $L$ at $4\degr$ to reduce the eye's
preference for distant stars and, also, so that the lines don't become
unrecognizably long. The assembly of lines is like iron filings near a
bar magnet and traces out the plane-of-the-sky field lines. Note that
these lines aren't vectors, because they don't indicate direction;
linear polarization is defined only modulo $180\degr$, not $360\degr$,
so it only has an orientation.

Figure~\ref{lbpol} shows the major large-scale features in the magnetic
sky: \begin{enumerate}

\item In the Galactic plane, the lines tend to be parallel to the plane,
showing that the large-scale field lies in the plane. This is expected,
if only from the effects of differential rotation and flux freezing.

\item Near $\ell = (80\degr, 260\degr)$ the lines lose this tendency.
Heiles (1996a) used this observed effect to determine the direction and
curvature of the local magnetic field: it points towards $\ell \sim
83\degr \pm 4.1\degr$ and has radius of curvature $8.8 \pm 1.8$\,kpc.

\item Figure~\ref{lbpol} shows several small areas where the density of
measurements is so high as to obliterate the individual lines. These are
regions of particular interest because of their dense clouds or star
formation. We label Orion, Taurus, and Perseus, but several others also
stand out. In these regions the dense clouds often look filamentary.

\hspace{5.0mm}  The observed stellar polarizations sometimes exhibit
good alignment with filamentary structures, but the sense of alignment
is not always the same. Three particularly good examples are
Pereyra \& Magalhaes (2004) and Fig.~5 in Heyer et al. (1987),
where the polarizations are strikingly perpendicular to the long axis
of the filaments, and Plate IX in Vrba et al. (1976), where the
polarizations are parallel. The proper interpretation of these
completely orthogonal senses of alignment {\em probably} consists of
the following: \begin{enumerate}

\item Interstellar ``filaments'' are edge-on sheets.

\item Molecular clouds are flattened triaxial ellipsoids, which
are often flattened enough to be considered as slabs (Sect.~\ref{basu}
below).

\item Fat interstellar filaments are the projections of flattened
ellipsoids at random angles onto the plane of the sky.

\item The apparent orientation of $\Bperp$ for such ellipsoids
can adopt any position angle (call it $\Psi$) because of projection
effects, as emphasized in the very important article by Basu (2000).

\item The only reliable way to determine the orientation of field lines
with respect to the flattened ellipsoids is to compare the observed
histogram $\Psi$ for a large sample with model probability
distributions for $\Psi$, such as Basu's. Not enough regions have been
measured to accumulate sufficiently large-number statistics on $\Psi$.
In particular, we caution that statements like ``the observed $\Bperp$
is perpendicular to the filament, i.e.\ perpendicular to the edge-on
sheet'' can be misleading when applied to a single example and can only
have validity when applied to a good statistical sample.

\end{enumerate}

\item Figure~\ref{lbpol} shows the prominent distortion of the local
field produced by Loop~I (also known as the North Polar Spur).  This
distortion is also visible in the $\HI$ line and radio synchrotron
continuum.  It is the result of a superbubble produced by stellar winds
and supernovae in the Sco/Cen association; the overall morphology of the
$\HI$, hot gas (from its X-ray emission), and magnetic field (from
radio synchrotron emission) strikingly confirms the concept that the
ISM is shaped by such explosions.  The center of Loop~I appears in
different places for the radio continuum (near $(\ell, b) \sim
(329\degr,18\degr)$ (Berkhuijsen et al. 1971) and for the
$\HI$ (near $(320\degr, 5\degr)$ (Heiles 1998b).
The causes for this difference are not currently understood.

\hspace{5.0mm} Note our discussion of the field distortion by
superbubbles in Sect.~\ref{magneticem}. The case here, with Loop~I,
is clear-cut because the ambient field lies predominantly across the
line of sight. Other geometries are less clear and more complicated.

\item There are other large scale patterns in Fig.~\ref{lbpol}, which
presumably trace other supernova shells or supershells. There is ample
opportunity for further research here! \end{enumerate}

\subsection{Polarization of Thermal Grain Emission} \label{grainemission}

Starlight polarization occupies a high position, not only because of its
historical importance but also because stars serve as distance markers.
However, as with any tracer dependent on background sources, it is not
very useful for mapping.  Thermal radiation from dust is polarized,
again because of the alignment of dust grains.  We can look forward to
the day when (1) enough stellar extinction measurements exist to
determine the evolution of extinction with distance along arbitrary
lines of sight, and (2) the mapping of IR emission from the diffuse
interstellar gas starts in earnest.  Unfortunately, (1) is in its
infancy, except for particularly well defined clouds of high extinction,
and regarding (2) no IR polarization data exist at all for diffuse
regions.

In dense regions, however, far-infrared and millimeter wavelength
observations of linearly polarized dust emission may be used to map the
morphology of the magnetic field projected onto the plane of the sky,
$\Bperp$ (Hildebrand 1988).  The position angle of maximum emission will
be perpendicular to $\Bperp$.  The mm-wavelengths sample the larger
aligned grains and have the advantage that local star formation is not
required because mm-wavelength emission occurs even with cold grains.
These are particularly useful for places where stars have formed,
because they heat the dust and provide strong emission.  These regions
are discussed later in this review.  Other recent reviews which cover
these aspects very well are Hildebrand et al.\  (2000), Hildebrand
(2002), and Crutcher et al.\  (2003).

It is not possible to measure directly the strength of $\Bperp$ since
fairly weak magnetic fields can align grains, so the degree of
polarization is not a measure of field strength.  However, in the early
days of interstellar polarization studies, Chandrasekhar and Fermi
(1953) suggested that analysis of the small-scale randomness of magnetic
field lines could yield estimates of the field strengths.  The method
depends on the fact that turbulent motions will lead to irregular
magnetic fields (since under interstellar conditions fields will be
frozen into the matter).  There will therefore be a perturbed or
MHD-wave component to the field that should show up as an irregular
scatter in polarization position angles relative to those that would be
produced by a regular magnetic field.  The stronger the regular field,
the more it resists being irregularized by turbulence.  They showed that
the magnitude of the irregularity of field lines could yield the regular
field strength in the plane of the sky: \begin{equation} \Bperp =
Q\sqrt{4\pi \rho} \; \frac{\delta V}{\delta \phi} \approx 9.3
\sqrt{n(H_2)} \;\frac{\Delta V}{\delta \phi} \; \mu G, \end{equation}
where $\rho = m n(H_2)$ is the gas density, $\delta V$ is the velocity
dispersion, $\delta \phi$ is the dispersion in polarization position
angles in degrees, $Q$ is a factor of order unity, $n(H_2)$ is the
molecular hydrogen density in molecules cm$^{-3}$, and $\Delta V =
\sqrt{8 ln2} \; \delta V$ is the FWHM line width in km~s$^{-1}$.  Here
we have used $Q = 0.5$, a calibration based on study of simulations of
interstellar clouds by Ostriker, Stone, and Gammie (2001), but see also
Heitsch et al.  (2001) and Padoan et al.  (2001).  These simulations
found that this method could yield reliable results in molecular clouds
if $\delta \phi < 25^\circ$.  One should note that while fluctuations in
the field along the line of sight will be smoothed out by the
polarization measurements, the calibration by the simulations referred
to above include this in the $Q$ factor.  Heitsch et al.  (2001) studied
the effects of smoothing due to inadequate spatial resolution in the
plane of the sky; although such smoothing will pro duce too large an
estimate of $\Bperp$, the problem can be overcome so long as the region
being studied, i.e.  a molecular cloud or core, is adequately (a few
resolution elements) resolved.  The Chandrasekhar-Fermi method of
estimating $B$ is a statistical one that may be in error by $\sim 2$ for
an individual cloud.

\subsection{Spectral-line linear polarization}

Linear polarization may also arise in radio-frequency spectral lines
formed in the interstellar medium, even when Zeeman splitting is
negligible.  This Goldreich-Kylafis effect (Goldreich and Kylafis 1981,
Kylafis 1983) may be used to probe magnetic field morphologies in
molecular clouds.  Heiles et al.  (1993) provide a qualitative
discussion of how the linear polarization arises.  The direction of the
polarization can be either parallel or perpendicular to the magnetic
field, depending on the relationship between the line of sight, the
direction of the magnetic field, and the direction of a velocity
gradient that produces the anisotropic line optical depth that is
required to produce linear polarization.  Although the theory makes
specific predictions for whether the field is parallel or perpendicular
to the line polarization, in general the observations do not provide all
of the necessary information.  This ambiguity is unfortunate, but if
structure in a cloud causes a flip by $90^\circ$ in the polarization
direction, it would easily be recognized and not confused with random
magnetic fields.  It therefore is a valuable tool in the measurement of
magnetic field direction and in the degree of randomness of the field.
As is the case for dust polarization, the Chandrasekhar-Fermi method may
be applied to maps of spectral-line linear polarization to estimate
field strengths.

\subsection{Zeeman Splitting} \label{zeemansplitting}

Interstellar magnetic fields are very weak and in all cases except
masers produce Zeeman splitting $\varDelta \nu_\mathrm{Z}$ that is much
smaller than the line width $\delta \nu$, so we usually have ${\varDelta
\nu_\mathrm{Z} / \delta \nu} \ll 1$.  This makes Zeeman splitting
observations sensitivity limited. Accordingly, the only hope of
detecting the splitting is with an atom or molecule whose splitting is
``large'', i.e.\ $\sim$ the Bohr magneton ${e {\bar h} / 2
m_\mathrm{e} c}$; this, in turn, means that the molecule must have a
large magnetic moment $\mu$ and Land\'e factor $g$.  Thus, only species
with electronic angular momentum are useful for Zeeman splitting
observations. Other molecules have splitting $\sim$ the nuclear
magneton ${e {\bar h} / 2 m_\mathrm{n} c}$, which is thousands of
times smaller. There is one spectacular exception, water masers, where
$\Bobs$ is tens of mG in regions having volume density $n \simgt
10^8\cmcube$ (Sarma et al. 2002).

For a given $\Bobs$, the splitting $\varDelta \nu_\mathrm{Z}$ depends on
$g$ but is independent of the line frequency itself.  For species with
higher line frequencies, the line widths $\delta \nu$ rise
proportionally, so for a given field strength the ratio ${\varDelta
\nu_\mathrm{Z} / \delta \nu}$ decreases proportionally. This ratio
is the crucial one for sensitivity, so in the absence of other
considerations it is better to use low-frequency spectral lines.
Heiles et al. (1993) describe the details and provide a list of atoms
and molecules having electronic angular momentum. Suitable low-frequency
($< 11.2$\,GHz) species include $\HI$, Radio Recombination Lines, OH,
CH, C$_4$H, and C$_2$S.  Other molecules have much higher frequencies,
but experience shows that this is not always devastating because they
can exist in very dense regions where field strengths are high enough to
compensate; the defining example is CN (Crutcher et al. 1999), with
line frequency $\sim 114$\,GHz and $\Bobs$ of several hundred $\umu$G
in the Orion Molecular Cloud~1, two cores in DR21OH, and probably
M17SW.

Although the Stokes parameters V, Q, and U for the Zeeman components
provide in principle full information about magnetic field strength and
direction, in practice full information on {\bf B} cannot be obtained
owing to the extreme weakness of Q and U. For the usual small-splitting
case ${\varDelta \nu_\mathrm{Z} / \delta \nu} \ll 1$, Zeeman splitting
is detectable in the Stokes $V$ spectrum, which is the difference
between the two circular polarizations.  The $V$ spectrum has the shape
of the first derivative of the line profile (the Stokes $I$ spectrum)
with an amplitude $\propto {B_{||} / \delta \nu}$, where $B_{||}$ is the
line-of-sight component of the field.

Why $\Bpar$ instead of $\Btot$? Or, in colloquial terms, how do the
interstellar atoms ``know'' where the observer is by arranging the
splitting to reveal only the particular field component that is oriented
towards the observer? The answer involves the directionality associated
with the circularly polarized line intensity.  In contrast, when
${\varDelta \nu_\mathrm{Z} / \delta \nu} > 1$ the observed effect is the
full splitting $\varDelta \nu_\mathrm{Z}$, which is $\propto \Btot$, not
$B_{||}$.  Crutcher et al.  (1993) treat this question in detail and
provide formulas for the general case.

\begin{figure}[!h]  \centering  \includegraphics[width=3.5in]
{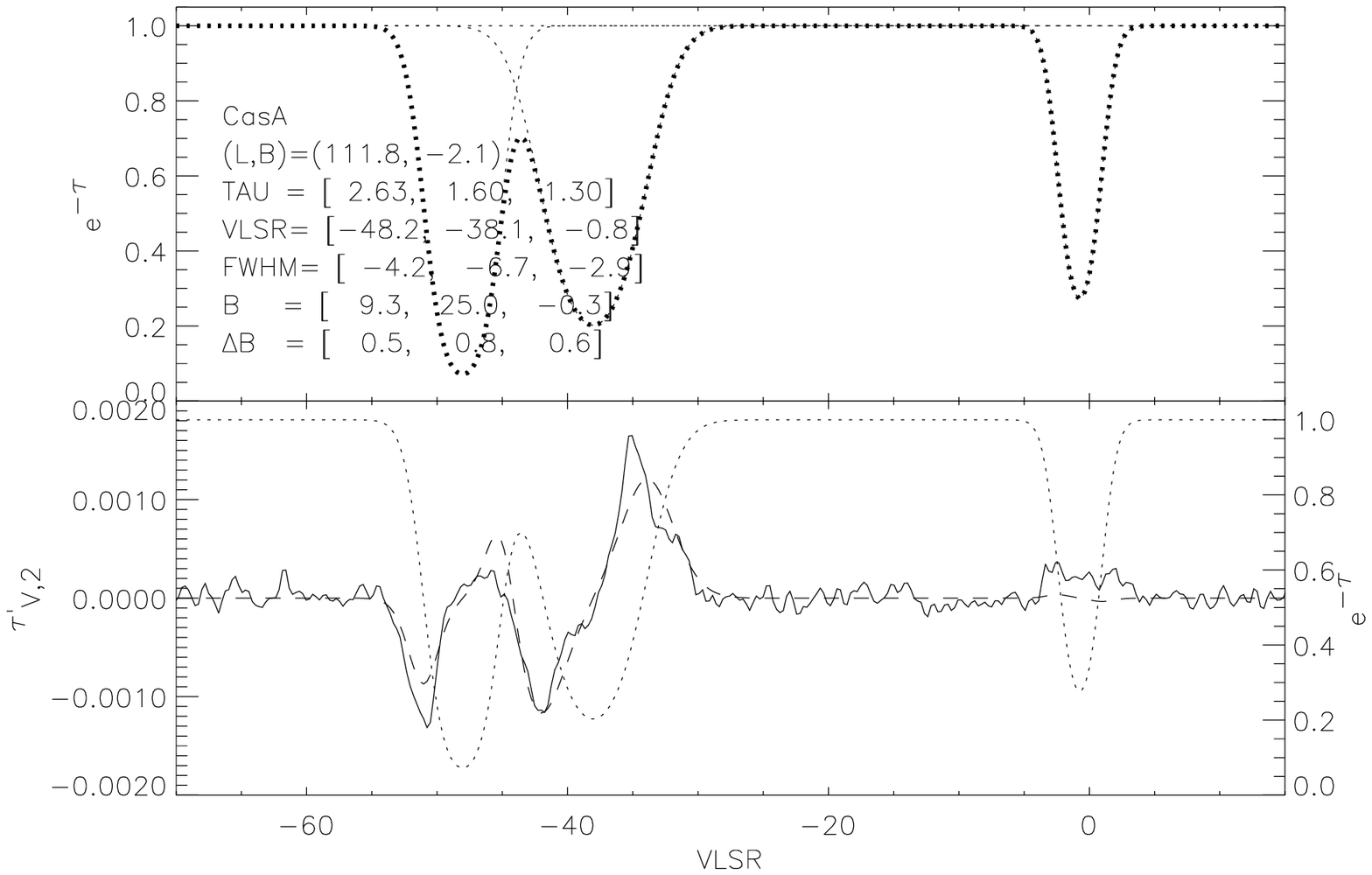} \includegraphics[width=3.5in] {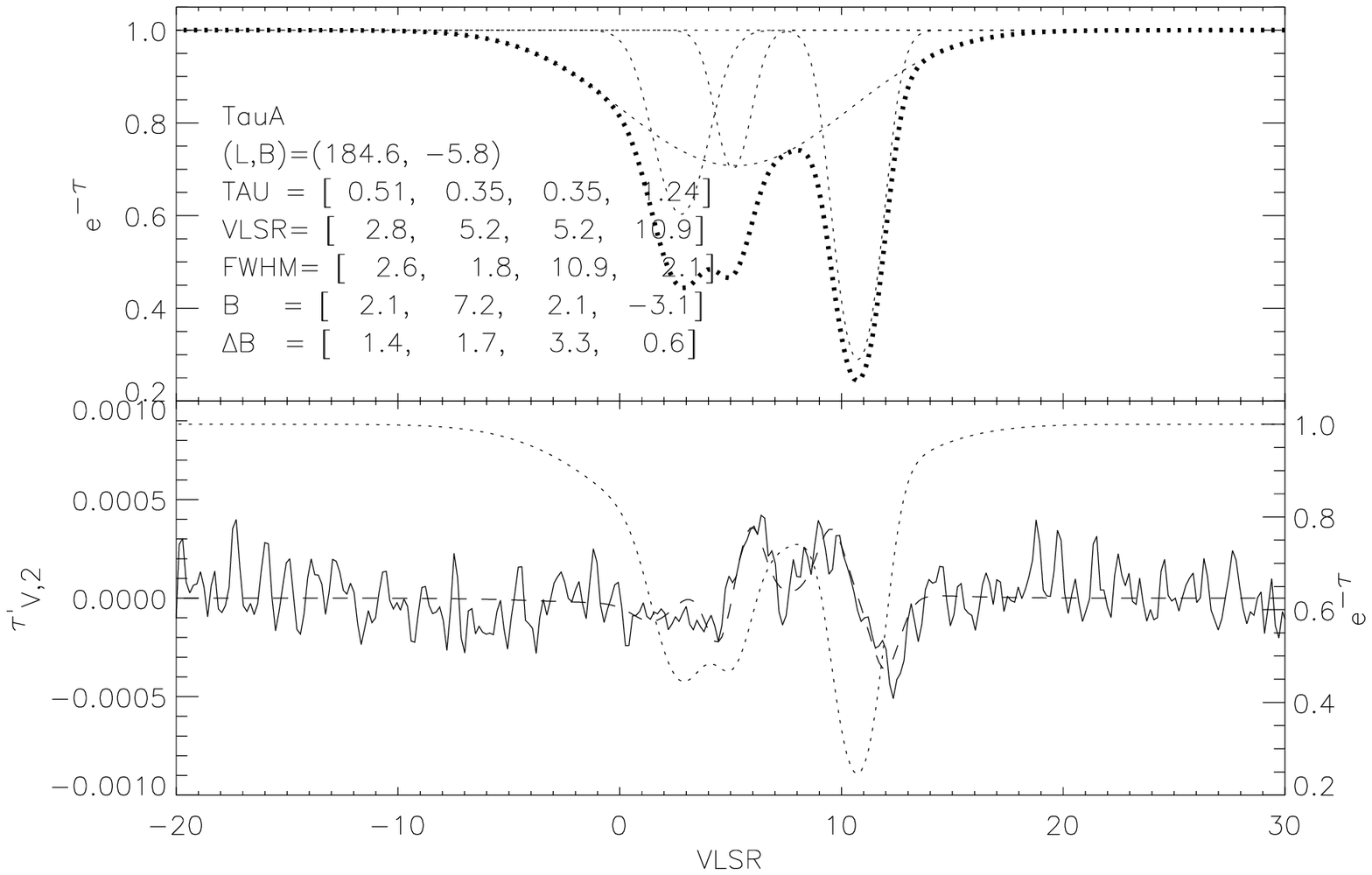}
\caption{Examples of $\HI$ Zeeman splitting for two sources in
absorption from Heiles \& Troland (2004).
The {\bf top panel} shows Cas~A (data from HCRO).
The {\bf bottom panel} is Tau~A. These are detections with very high
signal/noise.  See Sect.~\ref{millennium} for details. \label{zmnega} }
\end{figure}

\begin{figure}[!h]
\centering
\includegraphics[width=3.5in] {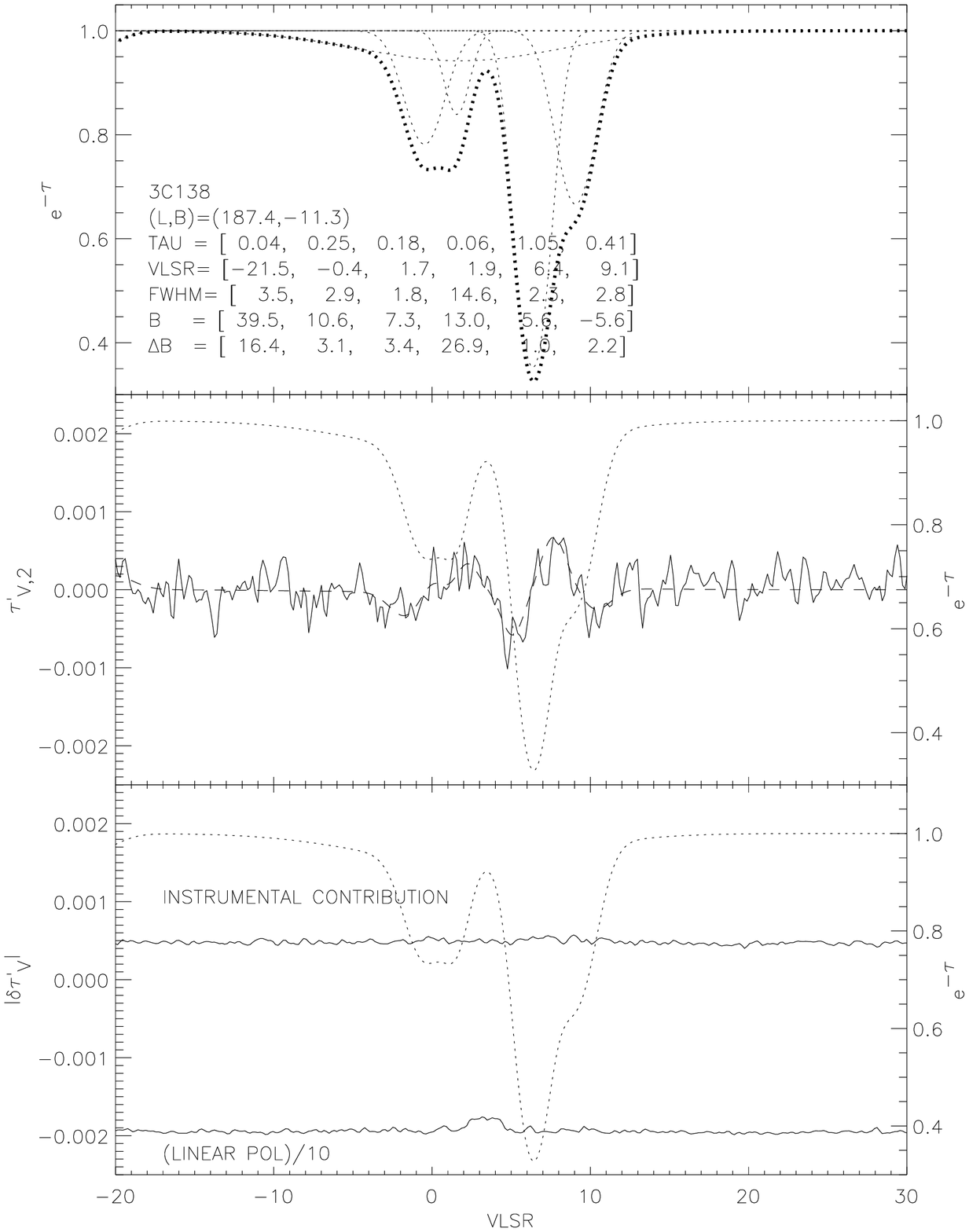}
\caption{An example of $\HI$ Zeeman splitting for 3C138 (Heiles \&
Troland 2004).
This measurement has high signal/noise relative to most other results
in the Millennium survey.  See Sect.~\ref{millennium} for details.
\label{zmnegb} }
\end{figure}

As examples of Zeeman splitting detections, Figs.~\ref{zmnega} and
\ref{zmnegb} illustrate Zeeman splitting for three sources from the
Arecibo Millennium survey (Heiles \& Troland 2004) in order of
decreasing signal/noise.  The top panel of Figure \ref{zmnega} shows
Cas~A [data from Hat Creek (HCRO)], with more than 100 hours of
integration, and the  bottom one shows Tau~A (from Arecibo), with $\sim
7$ hours. Figure \ref{zmnegb} shows 3C138 (from Arecibo) with $\sim 17$
hours.  See Sect.~\ref{binabs} for discussion. Fig.~\ref{DR21OH} shows a
molecular Zeeman detection for the 3-mm CN lines toward DR 21 (OH), and
Figs.~\ref{S106profiles} and \ref{S106maps} show a molecular Zeeman detection
and $\Bpar$ map for the 18-cm line of OH toward S 106.

\section{Observed vs.\ Intrinsic Probability Density Functions} \label{pdfs}

We begin our focus on data and their interpretation with a rather
technical discussion of the probability density function (pdf) of
observed components of magnetic field and how they relate to the total
field strength.  This turns out to be surprisingly important, and
because this discussion has not appeared prominently in past literature
we devote considerable attention to it.

\subsection{Conversion of the Intrinsic \boldmath{$\phi(\Btot)$} to the
Observed \boldmath{$\psi(\Bobs)$} and \boldmath{$\psi(\Bperp)$}}
\label{univarb}

Given a field strength $\Btot$ which can be randomly oriented to the
line of sight, what is the probability of finding an observed field
strength $\Bobs$? Alternatively, this is equivalent to the simple case in
which all clouds have the same $\Btot$, which is randomly oriented with
respect to the observer.  The line-of-sight component $\Bobs$ is

\begin{equation}
\Bobs = \Btot \cos \theta \ ,
\end{equation}

\noindent where $\theta$ is the angle between the field direction and
the line of sight. $\theta$ can run from 0 to $\pi$, but it's simpler
and no less general to consider the smaller interval $\theta$ from 0 to
$\pi / 2$. In this case, the pdf of $\theta$ is  the familiar

\begin{equation}
\phi_\theta(\theta) = \sin \theta
\end{equation}

\noindent and we wish to know the pdf of $\Bobs$, which is given by (see
Trumpler \& Weaver (1953) for a discussion of these conversions)

\begin{equation}
\psi({\Bobs}) = \phi_\theta[ \theta(\Bobs)] \left|
        {d[\theta(\Bobs)] \over d\Bobs} \right|  \ ,
\end{equation}

\noindent which gives

\begin{eqnarray} \label{Bobs}
\psi({\Bobs}) =
\left\{
\begin{array}{ll}
{1 \over \Btot} & {\rm if} \  0  \leq \Bobs \leq \Btot \\
        0       & {\rm otherwise}
\end{array}
\right. \ .
\end{eqnarray}

\noindent In other words, $\Bobs$ is uniformly distributed between the
maximum possible extremes 0 and $\Btot$ (actually $\pm\Btot$). This
leads to the well-known results that in a large statistical sample, both
the median and the mean observed field strengths are half the total
field strength and also $\Bpar^2 = {\Btot^2 / 3}$. More generally,
observed fields are always smaller than the actual total fields, and
with significant probability they range all the way down to zero.

Similarly, we can derive the pdf for $\Bperp$, the plane-of-the
sky component; this is important for starlight polarization and
synchrotron emissivity. We have

\begin{eqnarray} \label{Bperp}
\psi(\Bperp) =
\left\{
\begin{array}{ll}
{\Bperp \over {\Btot}^2 } \left[ 1 -
 \left( \Bperp \over \Btot \right)^2 \right]^{-1/2}
 & {\rm if} \  0  \leq \Bperp \leq \Btot \\
        0       & {\rm otherwise}
\end{array}
\right. \ .
\end{eqnarray}

\noindent The pdf $\psi \rightarrow \infty$ as $\Bperp \rightarrow
\Btot$, but the cumulative distribution is well defined. The mean and
median are $0.79\,\Btot$ and $0.87\,\Btot$, respectively; the high
values reflect the large fraction of slabs tilted to the line of sight,
where $\Bperp$ is large. The mean of ${\Bperp}^2$ is ${2/3}
{\Btot}^2$.

    The above applies if all $\Btot$ are the same. Now suppose
$\Btot$ has an arbitrary pdf $\phi(\Btot)$. Again, following standard
techniques, we obtain

\begin{equation} \label{Bpsiphi}
\psi(\Bpar) = \int_{[{\Bobs}>{\Btot}_{min}]}^\infty {\phi(\Btot)
        \over \Btot} d\Btot \ ,
\end{equation}

\noindent where the symbol $[{\Bobs}>{\Btot}_\mathrm{min}]$ means the
larger of the two quantities. The presence of $\Btot$ in the
denominator means that smaller ranges of $\Bobs$ are emphasized. This
is an obvious consequence of (\ref{Bobs})'s uniform pdf for a single
field value.

    Similarly, for $\Bperp$ we obtain the more complicated

\begin{equation}
\label{volterra1}
\psi(\Bperp) = \int_{\Bperp}^\infty
{\Bperp \over {\Btot}^2 }
\left[ 1 - \left( \Bperp \over \Btot \right)^2 \right]^{-1/2}
\phi(\Btot )  d\Btot \ .
\end{equation}

\begin{figure}[h!]
\begin{center}
\includegraphics[width=3.0in] {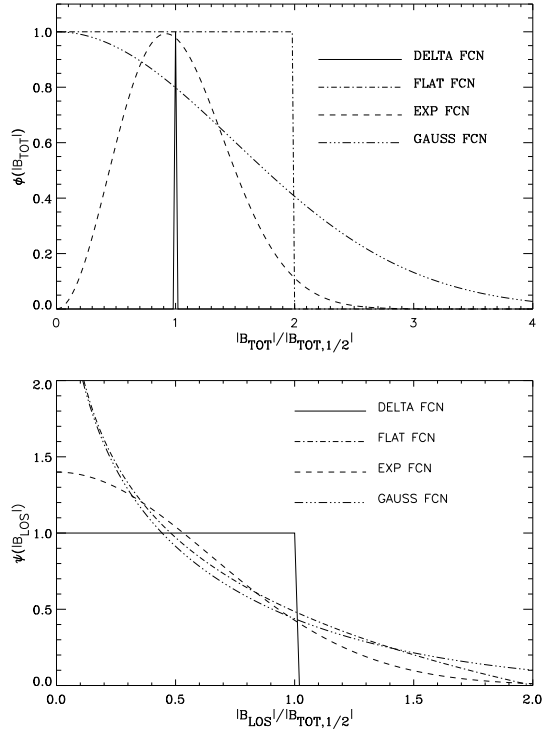}
\end{center}

\caption{Top panel: The intrinsic $\phi(B_{tot})$ for four
representative functional forms. Bottom panel: their line-of-sight
counterparts $\psi(B_{los})$. The vertical scales are arbitrary.
\label{pdf_fig0} } \end{figure}

    It's worth illustrating these equations with some examples.
Figure \ref{pdf_fig0} illustrates the solution of equation
\ref{Bpsiphi} for four functional forms of $\phi(B_{tot})$ plotted
against $|B| \over |B_{1/2}|$, where the subscript $1/2$ denotes the
median value. These forms include the following:
\begin{enumerate}

\item $\phi(\Btot )$ a Kronecker delta function (DELTA FCN),
$\phi(\Btot) = \delta(\Btot - {\Btot}_{,0})$, yielding $\psi$ a
flat function (as discussed immediately above, equation \ref{Bobs});

\item $\phi(\Btot )$ a flat distribution (FLAT FCN) between $0 \le |B_{tot}| \le
B_0$, yielding $\psi \propto \ln\left({ B_0 \over B_{los}}\right)$;

\item  $\phi(\Btot )$ a weighted Gaussian (EXP FCN),

\begin{equation} \label{Btotexp}
\phi({\Btot}) = {{\sqrt{2 \over \pi B_0^2}}} \ {\Btot ^2 \over 2B_0^2}
\  e^{-( \Btot ^2 / 2 B_0^2)} \ ,
\end{equation}

\noindent yielding $\psi$ a Gaussian with dispersion $B_0$.

\item $\phi(\Btot )$ a Gaussian (GAUSS FCN) with dispersion $B_0$, yielding
$\psi \propto E_1 \left(B_{los}^2 \over 2B_0^2 \right)$, where $E_1$ is
the exponential integral of order 1.

\end{enumerate}

        All four $\phi(B_{tot})$ are plotted with respect to $B_{tot}
\over B_{tot,1/2}$, so the medians of all lie at unity on the $x$-axis.
However, the means differ.  Similarly, the medians and means of the
associated $\psi(B_{los})$ differ from each other.  These relationships
between median and mean are summarized in Table \ref{medianmean}.  The
medians and means for $\psi(B_{los})$ are all about half those for
$\phi(\Btot )$, which is a direct result of the weighting by
$B_{tot}^{-1}$ in equation \ref{Bpsiphi}.

\begin{table}
\begin{centering}
\caption{MEDIANS AND MEANS OF FOR
REPRESENTATIVE PDFS  \label{medianmean} }
\hspace{1.0in}
\begin{tabular}{ccccc}
$\phi(B_{tot})$ & $B_{tot,1/2}$ & $\langle B_{tot} \rangle$ &
    $B_{los,1/2}$ & $\langle B_{los} \rangle$ \\
\hline
DELTA FCN & 1.00 & 1.00  & 0.50  & 0.50  \\
FLAT FCN & 1.00  & 1.00  & 0.40  & 0.52  \\
GAUSS FCN & 1.00 & 1.18  & 0.38  & 0.59  \\
EXP FCN   & 1.00 & 1.04  & 0.44  & 0.51  \\
\end{tabular}
\end{centering}
\end{table}

        Figure \ref{pdf_fig0} is disappointing from the observer's
standpoint, because the observed distributions $\psi(B_{los})$ do not
differ very much. These differences become smaller---inconsequential, in
fact---when there is some measurement noise. Unfortunately, given the
inevitable errors in {\it any} observation that is sensitive to
$B_{los}$, it seems practically impossible to distinguish among
different functional forms for $\phi(B_{tot})$. Nevertheless, the
average value of $B_{los}$ is close to half the average value of
$B_{tot}$ for a wide range of intrinsic pdfs of the latter; this also
applies to the medians, but less accurately. Therefore, this rule of
thumb may be used to estimate the median or average $B_{tot}$ from an
ensemble of measurements of $B_{los}$.


\subsection{Conversion of the Intrinsic \boldmath{$\phi[\log(\Btot)]$} to
\boldmath{$\psi[\log(\Bobs)]$}} \label{univarlogb}

Sometimes people treat $\log(\Bobs)$, instead of $\Bobs$, as the
important quantity. In particular, in Sect.~\ref{basu} below, we
consider least square fits of $\log(\Bobs)$ for molecular clouds. The
statistics for $\log(\Bobs)$ differ from those of $\Bobs$. Carrying
through the usual analysis, we find for the analog to (\ref{Bobs})

\begin{equation}
\psi \left[ \log \left( {\Bobs \over \Btot} \right) \right] =
\ln(10) \ 10^{\log \left( {\Bobs \over \Btot} \right) } \ .
\end{equation}

\noindent The mean and median of $\left[ \log \left( {\Bobs / \Btot}
\right) \right]$ are $-0.434$ and $-0.693$, which correspond to ${\Bobs
/ \Btot} = 0.37$ and $0.21$, respectively. Thus the statistics of
$\log \Btot$ favor smaller means and medians than do those of $\Btot$,
for which both numbers are 0.5.

\subsection{Conversion of the Intrinsic \boldmath{$\phi(\Nperp)$} to
the Observed \boldmath{$\psi(\Nobs)$} for Sheets} \label{univarN}

Many interstellar morphological structures are sheets. Examples for
$\HI$ include two sheets mapped in 21-cm line emission (Heiles 1967),
and an extreme sheet with aspect ratio of several hundred (Heiles \&
Troland 2003).
Along with Heiles \& Troland (2003), we consider that all CNM structures
are best considered as sheets.

As we did with the field, we discuss the pdfs of the observed column
density for sheets ($\Nobs$) given the total $\HI$ column density
$\Nperp$ in the direction perpendicular to the sheet, again assuming
random orientations. If the normal vector to the sheet is oriented at
angle $\theta$ with respect to the line of sight, then we have

\begin{equation}
\Nobs = { \Nperp \over \cos\theta} \ .
\end{equation}

\noindent If all sheets have the same $\Nperp$, then

\begin{eqnarray} \label{Nobs}
\psi(\Nobs)=
\left\{
\begin{array}{ll}
{\Nperp \over \Nobs^2} & {\rm if} \  \Nobs \geq \Nperp \\
        0       & {\rm otherwise}
\end{array}
\right. \ .
\end{eqnarray}

\noindent For a single $\Nperp$, $\Nobs$ has a long tail extending to
infinity. The mean value of $\Nobs$ is not defined because, with
infinite sheets, the integral diverges logarithmically; of course, this
doesn't occur in the real world, where sheets don't extend to infinity.
The median value of $\Nobs$ is $2 \Nperp$, reflecting the increased
observed column for tilted sheets. For an arbitrary pdf $\phi(\Nperp)$
we obtain

\begin{equation}  \label{Npsiphi}
\psi(\Nobs )= {1 \over \Nobs ^2} \int_0^{[\Nobs <
{\Nperp}_\mathrm{max}]}
        \Nperp \ \phi(\Nperp) \ d\Nperp \ .
\end{equation}

\subsection{Conversion of the Intrinsic Bivariate Distribution
\boldmath{$\phi(\Btot, \Nperp$)} to the Observed \boldmath{$\psi(\Bpar,
\Nobs)$} for Sheets} \label{bivarBN}

We can reasonably expect the magnetic field to lie either parallel or
perpendicular to the sheet. If the sheet has formed by coalescence of
more diffuse gas flowing more easily along the field lines, then the
field should lie perpendicular to the sheet. In contrast, if the sheet
is the result of a shock that has swept up both the gas and magnetic
field lines, then the field should lie parallel to the sheet.
Accordingly, we are led to consider the bivariate distribution of
magnetic field and column density for these two cases. We  assume that
$\Btot$, $\Nperp$, and of course $\theta$ are all uncorrelated. We again
consider the illustrative case of delta functions for $\Btot$ and
$\Nperp$.

\begin{figure}[!h]
\centering
\includegraphics[width=3.5in] {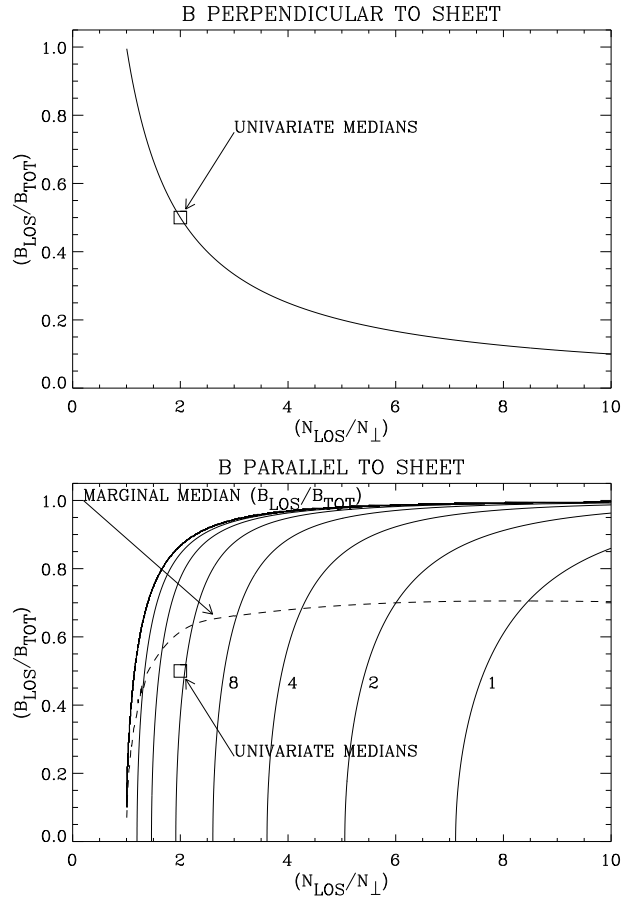}
\caption{The theoretical observed joint pdfs $\psi(\Bpar, \Nobs)$ for
the illustrative case of $\delta$-function distributions for $\Bpar$ and
$\Nobs$. The {\bf top panel} shows the pdf for $\Btot$ perpendicular to
the sheets; it degenerates into a single line. The {\bf bottom panel} is
for $\Btot$ parallel to the sheets; contours are spaced by factors of 2
with arbitrary scaling, and the dashed line shows the median $\Bobs$
versus $\Nobs$. \label{twodpdfplot} }
\end{figure}

    If $\Btot$ is perpendicular to the sheet (the {\em perpendicular
model}), then both $\Bobs$ and $\Nperp$ depend only on $\cos \theta$, so
the bivariate pdf degenerates to the deterministic line

\begin{equation}
\Bobs= {\Btot} {{\Nperp} \over \Nobs}
\end{equation}

\noindent which is shown in the top panel of Fig.~\ref{twodpdfplot}.
The parallel model, with $\Btot$ lying in the sheet, is more
complicated, with

\begin{equation}
\psi(\Bpar , \Nobs)=
{ {\Nperp} \over \pi \Nobs}
[ ({\Btot} \Nobs)^2 - ({\Btot} {\Nperp})^2 -(\Bpar \Nobs)^2 ]^ {-1/2}
\ .
\end{equation}

\noindent This is illustrated by the contours in the bottom panel of
Fig.~\ref{twodpdfplot}.

\subsubsection {Discussion of Figure~\ref{twodpdfplot}}
\label{twodpdfsection}

The two panels of Fig.~\ref{twodpdfplot} exhibit the joint pdfs for the
two sheet models ($\Btot$ perpendicular and parallel to the sheets).
The median observed column density ${\Nobs}_{1/2}$ is twice the assumed
$\Nperp$ and the median observed magnetic ${\Bpar}_{1/2}$ is half the
assumed $\Btot$; these univariate medians are indicated by squares on
the top two panels. The significance of these squares is that half the
observed $\Bpar$, and half the observed $\Nobs$, are smaller and half
larger. Finally, the dashed line in the middle panel exhibits the
median ${\Bpar}_{1/2}$ versus $\Nobs$; we calculate this by extracting
the conditional pdf $\psi(\Bpar | \Nobs)$ versus $\Nobs$, and
calculating the medians from its cumulative distributions.

The top and middle panels illustrate a crucial observational signature
at large $\Nobs$ that distinguishes between the two sheet models: for
the perpendicular model, large $\Nobs$ goes with small $\Bpar$, and
vice-versa for the parallel model. More quantitatively, for the
perpendicular model, {\em all} of the datapoints having $\Nobs$ above
its univariate median ($\Nobs > {\Nobs}_{1/2}$, indicated by the
square) have $\Bpar < {\Bpar}_{1/2}$. In contrast, for the parallel
model {\em most} ($66\%$) of the datapoints with $\Nobs >
{\Nobs}_{1/2}$ have $\Bpar > {\Bpar}_{1/2}$. More precisely for the
parallel model, as $\Nobs$ gets large, the marginal pdf $\psi( \Bpar \,
| \, \Nobs) \rightarrow {{\Nperp}_0 / \pi \Nobs^2}\, ({\Btot}_0^2 -
{\Bpar}^2)^{-1/2}$, which produces the median ${\Bpar}_{1/2} \rightarrow
0.71\, \Btot$; this is the asymptote of the dashed line on the middle
panel of Fig.~\ref{twodpdfplot}.

\subsection{Commentary} \label{pdfcommentary}

This discussion has been rather technical, more so than is usual in a
review.  However, the payoff follows because we can make some powerful
inferences from this discussion.  \begin{enumerate}

\item Consider the one-dimensional $\psi(\Bperp)$ for a given $\Btot$.
$\psi(\Bperp)$ diverges as $\Bperp \rightarrow \Btot$; the median and
mean values of $\Bperp$ are $0.79\, \Btot$ and $0.87\, \Btot$,
respectively. Thus, maps of starlight polarization, or IR polarization
of dust emission, tend to represent the full field strength to a
considerable degree, a much higher degree than does Zeeman splitting for
$\Bpar$ (see next paragraph).

\item Consider the one-dimensional $\psi(\Bpar)$ for a given $\Btot$.
$\psi(\Bpar)$ is flat for $0 < \Bpar < \Btot$.  Suppose we have a
collection of measured $\Bobs$ and can reasonably expect the
orientation to be random.  Suppose we wish to fit a dependence of
magnetic field on, say, volume density, as we will do below in
Sect.~\ref{basu}.  Then we should {\em not} use the standard least
squares technique because it assumes that the residuals from the mean
have a Gaussian distribution; in contrast, the intrinsic distribution of
residuals of $\Bobs$ is flat.  In particular, this means that errors
derived from the distribution of residuals to the fitted points are not
calculated correctly.

\hspace{5.0mm} Similarly, when fitting $\log \Bpar$ the distribution of
residuals is asymmetric, which introduces a systematic bias into the
least-squares fitted result. This must be corrected for, as we do in
Sect.~\ref{basu} below. In addition, of course, the errors
are also not calculated correctly.

\item Consider an assembly of $\Bobs$ from different sources, all of
which have the same $\Btot$.  Then we expect some $\Bobs$ to be very
small.  Thus, small values of $\Bobs$ do not necessarily mean that
$\Btot$ is small.  Rather, an unbiased survey produces many small,
undetectable values of $\Bobs$, which can be very frustrating for the
observer but is nevertheless inevitable.  A spectacular example is the
local-arm ($0\kms$) field seen against Cas~A (top panel
Fig.~\ref{zmnega}), $\Bobs = -0.3 \pm 0.6\,\umu$G.  This surprisingly
small result is perfectly consistent with statistical expectation.  Of
course, we cannot rule out that the field actually is really small in
any particular case like this, but one needs additional data to draw
such a conclusion!

\item Consider the large set of magnetic fields observed in 21-cm line
{\em emission} in morphologically obvious structures, reviewed below in
Sect.~\ref{magneticem}. The term ``morphologically obvious'' means
filaments or edge-on sheets.  Edge-on sheets should be edge-on shocks
in which the field is parallel to the sheet, i.e.\ with large $\theta$.
Here, the statistics reverse and favor relatively large $\Bpar$.  As
explained in Sect.~\ref{twodpdfsection}, as the line of sight
becomes parallel to the sheet -- i.e. for a morphologically obvious
sheet -- the median ${\Btot}_{1/2} \rightarrow 0.71\, \Btot$. For these
structures, measured fields are strong, ranging from $\sim 5$ to $\sim
10\,\umu$G.  This is not inconsistent with a uniform $\Btot \sim 10\,
\umu$G, which is a factor of two above the median CNM field strength
from Sect.~\ref{binabs}.  This suggests that shocks enhance the field
strength, but not by large factors.

\end{enumerate}

\section {\boldmath{$\Bobs$} from \boldmath{$\HI$} Absorption Lines}
\label{binabs}

Zeeman splitting of the $\HI$ line in absorption holds the enviable
position of being the means by which the interstellar magnetic field
{\em strength} was first measured (Verschuur 1969). With
quantitative knowledge of the magnetic field strengths came the
beginning of the end of the famous theorists' refuge (``\dots the
larger one's ignorance, the stronger the magnetic field'' (Woltjer
1967)).

Zeeman splitting in absorption, instead of emission, is enviable
for another important reason.  It is easier to measure $B_{||}$ in the
CNM than in the WNM because the $\HI$ line opacity $\propto T^{-1}$,
which makes the CNM appear prominently in absorption.  We detect
absorption by performing $(ON-OFF)$ measurements against a radio
continuum source; for such measurements the sidelobe contributions from
the emission tend to cancel. This makes the CNM absorption results
very much less subject to instrumental effects than emission results
(Heiles \& Troland 2004).
In fact, we consider the results to be statistically reliable, with
Gaussian-distributed uncertainties and small systematic errors.

\subsection{Early Work} \label{earlyw}

Verschuur's (1969) discovery of Zeeman splitting in interstellar $\HI$,
in absorption against against Cas~A and Tau~A, broke an earlier series
of frustrating efforts focused at Jodrell Bank\footnote{Verschuur made a
typographical error in labeling the sign of his Stokes $V$ profiles (but
not his derived $\Bobs$). In addition, higher sensitivity results
(Fig.~\ref{zmnega}; also Heiles \& Troland 2004) reveal more Gaussian
components with detected fields.}. He continued making such
measurements, but obtained physically interesting upper limits or
measurements for only five sources, which he reviewed in 1974 (Verschuur
1974). Four of these sources had detections.

Most of Verschuur's absorption detections do not refer to diffuse $\HI$,
but rather to molecular clouds or star-forming regions. Two of the four
sources (Orion~A and M17; $\Bpar \sim -60$ and $+25\,\umu$G
respectively) are dynamically active $\HII$ regions. One (two components
in the Cas~A Perseus arm with $\Bpar \sim (+9, +25)\,\umu$G;
Fig.~\ref{zmnega}) is a molecular cloud probably undergoing star
formation (Troland et al. 1985, Schwarz et al. 1986). None of these
refer to interstellar diffuse $\HI$. For sources that sample the diffuse
$\HI$, we are left with a single detection: Tau~A, with two velocity
components having $\Bobs \sim (-3, +7)\,\umu$G (Fig.~\ref{zmnega},
Heiles \& Troland 2004). Two other diffuse-cloud sources have only upper
limits:  Cygnus~A, with $\Bobs \simlt 3.5\,\umu$G, and Cas~A Orion arm,
with $\Bobs \simlt 1\,\umu$G (Fig.~\ref{zmnega}, Heiles \& Troland
2004).

Contrary to the usual development of observational astronomy,
Verschuur's discovery was not followed by the establishment of a
``cottage industry'' that produced a large number of detections
resulting in a significant expansion of $\HI$ absorption Zeeman
splitting measurements. The reason is simply the weakness of the Zeeman
splitting: typically ${\varDelta \nu_\mathrm{Z} / \delta \nu} \simlt
10^{-3}$. This state of affairs lasted until the turn of the millennium
(Heiles \& Troland 2004). 

\subsection{Recent Work: the Arecibo Millennium Survey} \label{millennium}

In our recent Arecibo Millennium survey, we (Heiles \& Troland 2004)
have only 22 detections that exceed $2.5\sigma$, out of a total of 69
measurements whose uncertainties are low enough to make them interesting.
This weakness forces us to discuss the CNM Zeeman splitting results
statistically.  And fortunately, the statistical reliability allows us
to actually carry through this statistical discussion.

Figures~\ref{zmnega} and \ref{zmnegb} exhibit three sources from the
Millennium survey as examples of strong detections. The top two panels
show Verschuur's original discovery sources Cas~A and Tau~A, but with
higher sensitivity than his original spectra. The separate detections
in two velocity components of the Perseus Arm, near $-40\kms$, are
very clear; the absence of a detection for the Orion arm near $0\kms$
is also clear. For Taurus, there are multiple Gaussian components, more
than one of which has associated features in Stokes $V$. The
multiple-component aspect is also clear for 3C138. For these sources
with multiple velocity components, we fit fields independently to each
component (Heiles \& Troland 2004).
The dashed lines in the three Stokes $V$ spectra show the fits.

We emphasize that these three sources have the strongest signal/noise
in Stokes $V$ in the entire sample. Mostly we obtain upper limits
instead of detections for $\Bobs$. When we include only those for
which the uncertainty $\varDelta \Bobs < 10\,\umu$G, the observed
histogram $\psi(\Bpar)$ resembles a Gaussian. Relating this to the intrinsic
field $\Btot$ is a complicated business requiring a Monte Carlo
analysis. The end result is that the median $\Btot$ is

\begin{equation}
{\Btot}_{,1/2} = 6.0 \pm 1.8 \ \mu {\rm G} \ .
\end{equation}

\noindent Not surprisingly from our earlier discussion, nothing can be
said about the pdf $\phi(\Btot)$.

\subsection{Equipartition Between Magnetism and
Turbulence in the CNM} \label{cnmequipartition}

There are no obvious correlations of $\Bobs$ with any quantity,
including $\Nobs$, linewidth, or $T_\mathrm{k}$.  However, we can
compare energy densities.

Each CNM component in Heiles \& Troland (2004)
is characterized by measured values of not only magnetic field but also
temperature, column density, and velocity dispersion. This allows us to
compare energy densities. One way to do this is with the classical
plasma parameter $\beta$, equal to the ratio of thermal to magnetic
pressure or, alternatively, thermal to magnetic energy density. We can
similarly define the ratio of turbulent to magnetic energy density
(Heiles \& Troland 2004).

For comparison of turbulent and magnetic effects in the CNM, we
calculate the relevant ratios for the following adopted parameter
values, which are close to the medians:

\begin{equation} \label{typicalvalues}
T = 50 \, {\rm K}
\end{equation}
\begin{equation}
\varDelta V_\mathrm{turb,1d} = 1.2 \kms
\end{equation}
\begin{equation}
\Btot = 6.0\,\umu{\rm G}
\end{equation}

\noindent These values provide

\begin{equation}
\beta_\mathrm{th} = 0.29
\end{equation}

\noindent and

\begin{equation} \label{equipartition}
\beta_\mathrm{turb} = {E_\mathrm{turb} \over E_\mathrm{mag}} =
M_\mathrm{ALF,turb}^2 = 1.3 \ .
\end{equation}

\noindent These values should be regarded as representative. Not all CNM
clouds have the median values, so these parameters have a considerable
spread.

\subsection{Field Strengths in the CNM Versus Those in Other Phases}

As mentioned in Sect.~\ref{intro}, Beck (2001)
reviews the most recent estimate of field strength derived from synchrotron
emission, minimum energy arguments, Faraday rotation, and polarization.
He finds the regular component to be $\sim 4\,\umu$G and the total
component to be $\sim 6\,\umu$G. The difference between regular and
total components is the fluctuating component, whose scale length is
probably at least tens of parsecs. Because our CNM structures are
physically small, it is more appropriate to compare their field
strengths with the total component. The CNM median of $\sim 6\umu$G
is nominally identical to Beck's local Galactic total component of $\sim
6\,\umu$G.

All of the other diffuse ISM phases are less dense than the CNM. For
example, both the WNM and the WIM are nearly two orders of magnitude
less dense. Thus the ISM field strength does not depend very sensitively
on volume density. In contrast, for the larger densities associated with
molecular clouds, in which gravity plays a significant role, the field
strength does increase with density, roughly $\Btot \propto n^{1/2}$
(Crutcher 1999).
The density independence for diffuse gas is well known from past
studies (Crutcher et al. 2003),
so this is hardly news; nevertheless, we tend to forget these things
and, moreover, from an observer's standpoint the paucity of detectable
fields is disappointing.

\subsection{Astrophilosophical Discussion} \label{apdisc}

These numbers indicate that magnetism and turbulence are in approximate
equipartition.  The approximate equipartition suggests that turbulence
and magnetism are intimately related by mutual exchange of energy.
Magnetic energies do not dissipate because the magnetic field cannot
decay on short time scales.  On the contrary, supersonic turbulence does
dissipate rapidly: numerical simulations of turbulence suggest that the
magnetic field does not mitigate turbulent dissipation (MacLow et al.
1998).  Thus, the equipartition between the dissipative turbulent energy
and nondissipative magnetic energy must arise from a mechanism other
than energy decay.

We suspect the answer is that the CNM components result from the
transient nature of turbulent flow: the CNM occupies regions where
densities are high, produced by converging flows, and the density rise
is limited by pressure forces.  This idea is discussed and reviewed
thoroughly by MacLow \& Klessen (2004).  These limiting pressures are
magnetic because the gas has small $\beta_\mathrm{th}$, meaning that
thermal pressure is negligible and the dynamical equality makes the
magnetic pressure comparable to the converging ram pressure.  The
equipartition looks like a steady-state equilibrium, but it is really a
snapshot of time-varying density fields and our immediate observational
view is a statistical result over a large sample.  In other words, our
current observational snapshot shows an ensemble at a given time.
Against this we compare the numerical simulations, which are stationary
in the sense that they have been allowed to run long enough that the
statistical properties become time-independent.  Such simulations are
also ergodic, with statistical properties over time being equivalent to
those over space.  With this view, the ISM dynamically evolves through
turbulence and its properties are governed by statistical equilibrium of
energy inputs and dissipation.

An alternative picture is based on the classical model of  {\em
static equilibrium} in which all forces balance.  Static clouds are
formed and evolve by gas moving adiabatically from one equilibrium
state to another as ambipolar diffusion allows magnetic flux to slowly
unfreeze. These slow adjustments in morphology occur primarily along
the field lines. At each stage there is a well-defined morphological
structure in quasistatic equilibrium. This idea was originated by
Mouschovias (1976) and has been well-developed by the
``Mouschovias school'' of students and collaborators, consisting of
Ciolek, Fiedler, and Basu (see Ciolek \& Basu 2000 and references
quoted therein), and by Shu and collaborators (see Shu et al. 1999).
The picture of static equilibrium predicts the linear
relationship between $\Btot$ and $\sigma_\mathrm{v} n^{1/2}$, which is
found for molecular clouds (Sect.~\ref{basu} below), which
is equivalent to the energy equipartition found in (\ref{equipartition}) above.

Both models predict the same result, namely approximate equipartition
between turbulent and magnetic energy densities. However, the concepts
on which they are based are in direct opposition. Which one is correct
for diffuse clouds? The role of gravity in diffuse clouds is
negligible. Given this, the static equilibrium models, for which gravity
is a major player, cannot apply to diffuse clouds. Thus, for diffuse gas
(but not for molecular clouds) we favor the concept of
statistical equilibrium as briefly outlined above. Analytical and
numerical research is being intently pursued on this topic; an excellent
review is MacLow \& Klessen (2004).

\section {\boldmath{$\Bobs$} from \boldmath{$\HI$} Emission Lines}
\label{bemission}

Zeeman splitting of the $\HI$ line in emission holds the enviable
position of not requiring a background source:  one can look anywhere,
so that the field in interesting regions can be measured and mapped.
However, nothing comes for free: emission measurements are prone to
instrumental error from polarized sidelobes. These errors have been the
subject of much controversy and here we will devote considerable
attention to explaining these matters. We will conclude that most
published Zeeman detections in $\HI$ emission are fairly reliable. We
begin our examination of this question with a discussion of instrumental
effects arising from polarized structure in the telescope beam.

\subsection{Instrumental Effects from Polarized Sidelobes and their
Description by a Taylor Series} \label{taylor}

The instrumental effects in $\HI$ Zeeman splitting measurements arise
from angular structure in the Stokes $V$ beam interacting with $\HI$
structure on the sky.  The $V$ beam has angular structure, even to the
extent of having sign changes. Troland \& Heiles (1982)
used both their empirical investigations of the HCRO telescope and
theoretical investigations published by others to classify this $V$
structure into three primary categories; here we split one, the sidelobe
component, into two subcomponents, near and far sidelobes. This gives:
\begin{enumerate}

\item {\em Beam squint}, in which the two circular polarizations point
in slightly different directions with typical separation
($\Psi_\mathrm{BS}$) of a few arcseconds.  This angular separation
doesn't seem like much, but given a small velocity gradient with
position the two beams see different frequencies, and this mimics the
tiny splitting resulting from the Zeeman effect.

\item {\em Beam squash}, in which the Stokes $V$ beam has slightly
different beamwidths in orthogonal directions.  These ``four-lobed''
polarized beams, in which two lobes on opposite sides of beam center
have the same sign and two lobes rotated $90\degr$ in position angle
have the opposite sign, are sometimes described as ``cloverleafs''.
This four-lobed structure responds to the second derivative of the 21-cm
line on the sky. Theoretically, beam squash occurs only for the linearly
polarized Stokes parameters $Q$ and $U$, but in practice it can also for
Stokes $V$ (e.g.\ Heiles et al. 2001, Heiles et al. 2003).

\item {\em Near-in sidelobes}, which can be considered as standard
diffraction effects and have polarization structure similar to that of
the main beam described above.

\item {\em Far-out sidelobes}.  For most telescopes the total power in
these ``distant sidelobes'' is nontrivial: even though the sidelobes are
weak, they cover very large solid angles and tend to be elliptically
polarized.  Troland and Heiles (1982) present one of the very few,
perhaps the only, map of the circular polarization of far-out sidelobes;
the pattern looks like a windmill and obviously results from feed legs.
These distant sidelobes are a result of telescope surface roughness and
the feed leg structure, so their structure is impossible to predict and
can be time variable.

\end{enumerate}

The classification is useful because it allows one to parameterize the
beam polarization effects. These parameters can be measured and
corrections applied. Nearly all $\HI$ emission Zeeman splitting
measurements have made these corrections in one form or another.

The appropriateness of this fourfold classification applies to all
telescopes that have been used for emission Zeeman splitting
observations: HCRO (Heiles 1996b),
the Green Bank 140-foot telescope (Verschuur 1969, 1989),
Arecibo (Heiles et al. 2001),
and the Green Bank Telescope (GBT) (Heiles et al. 2003).
For example, Verschuur's (1969) Fig.~2 presents the $V$ beam pattern
for the 140-foot telescope as it was in the late 1960's.  At that time,
it was very well described by beam squint with a peak-to-peak amplitude
of about 1.4\%; this corresponds to a beam squint $\Psi_\mathrm{BS}
\approx 7\arcsec$. Our maps of the complete polarized sidelobe
structure of the HCRO telescope always produced similar results,
although with much smaller beam squint.  Verschuur's (1989)
Fig.~1 presents the 140-foot polarized beam structure as it was in the
late 1980's, and shows a drastic difference: the newer map shows
primarily the four-lobed pattern of our category (2) with little beam
squint.  (The feed system had been changed between the two epochs.) The
1960's version of the beam pattern made the 140-foot
telescope unsuitable for Zeeman-splitting measurements of $\HI$ in
emission because the beam squint contribution to instrumental error
would have been excessive.  However, the 1980's version, with its small
beam squint but higher second-derivative component, was satisfactory --
as shown by the fact that Verschuur reobserved four positions that had
previously been observed with the HCRO telescope and found excellent
agreement in three.

\subsection{Verschuur's Bombshell} \label{bombshell}

Measurements of Zeeman splitting of $\HI$ emission lines have been made
by Troland, Heiles and other collaborators, and Verschuur.  Until 1993,
the agreement was quite good.

Despite the apparent agreement of the measurements, in 1993
Verschuur became highly suspicious of all emission results and dropped a
bombshell.  He asserted that ``\dots claims of Zeeman effect detections
in $\HI$ emission features \dots based on observations made with
presently available single-dish radio telescopes cannot be regarded as
reliable.'' At the time of his paper, the HCRO telescope had already
been destroyed, but he meant his claim to apply to that telescope as
well as other telescopes that were then available. This is a strong
statement and it has had a dampening effect on the field, making many
astronomers highly suspicious of the published results. Accordingly, we
believe a thorough discussion is in order. This discussion is excerpted
from Heiles (1998a), a reference which is difficult to find.

We believe Verschuur's claim to be incorrect. His claim is based on his
estimates of the instrumental effects, which in turn are based {\em
solely} on measurements of the velocity gradient of the $\HI$ line
(Verschuur 1995a,b).
In particular, his estimates of the instrumental effects are not based
at all on the {\em properties of the polarized beam}. To clarify his
procedure and its inadequacy, we describe its six steps:
\begin{enumerate}

\item Observe $V$ and $I$ spectra at the central position P; denote
these ${\rm V}_\mathrm{obs}(v)$ and $I_\mathrm{obs}(v)$.

\item Make an 8-point map of $I$ spectra around P. Each map position is
displaced from P by $15\arcmin$; in position angle the 8 points are
equally spaced ($45\degr$), with the displacements of 4 points towards
the cardinal directions in equatorial coordinates.

\item Find the pair of profiles whose difference spectrum $\varDelta
(v)$ is strongest and mimics the shape of V$_\mathrm{obs}(v)$.

\item Find the coefficient $R$ that scales the $\varDelta (v)$ spectrum to
the V$_\mathrm{obs}$ spectrum, i.e.  the best fit for $R \varDelta (v) =
{\rm V}_\mathrm{obs}(v)$.

\item Produce the ``corrected'' $V$ spectrum ${\rm V}_\mathrm{corr}(v) =
{\rm V}_\mathrm{obs}(v) - R \varDelta (v)$.

\item Derive the Zeeman splitting from ${\rm V}_\mathrm{corr}(v)$.
\end{enumerate}

The fatal flaw is that $R$, which represents the beam squint, is not
measured directly for the {\em telescope}. Rather, it is given the
particular value that minimizes the observed $V$ spectrum
$V_\mathrm{obs}(v)$.

As explained above, the beam squint samples the first derivative of the
21-cm line on the sky, which must contain a velocity gradient at some
level. Steps 2 and 3 of the above procedure measure the velocity
gradient. Step 4 fits this velocity gradient to the observed $V$
spectrum and derives the coefficient $R$. Then, no matter how large
$R$ is, it is used to subtract away the scaled $\varDelta$ profile from the
observed $V$ spectrum. With this step, $R$ implicitly represents the
amplitude of the beam squint in units of $30\arcmin$.

But the amplitude of the beam squint can be independently measured for
a telescope. The proper procedure would be to measure the beam squint
and velocity gradient, multiply the two vectorially, and subtract the
result from the observed $V$ spectrum.

Consider one particular entry in Verschuur's (1995b)
Table~2 as an example: NCPShell.4.  For this position he obtains $R =
0.0052$.  This corresponds to a beam squint of ($30\arcmin \times
0.0052) = 9\farcs 4$.  He uses this value of $R$ to subtract away a
velocity derivative from the V$_\mathrm{obs}$ profile that amounts to
$10.8\,\umu$G, obtaining a ``corrected'' field strength $2.1 \pm 1.0\,
\umu$G. In doing this he has removed the contribution to
V$_\mathrm{obs}$ that arises from the magnetic field -- he has removed
the ``signal''. In colloquial English, this is known as ``throwing out
the baby with the bathwater''.

The data in Verschuur's papers (1995a,b)
could be reanalyzed taking account of the fact that the beam squint of
the 140-foot telescope is limited to some maximum value. Unfortunately,
this is not discussed by Verschuur, but judging from his earlier paper
in this field (Verschuur 1989) the upper limit on 140-foot beam squint
is probably $\sim 3\arcsec$, which corresponds to $R = 0.0017$ (0.17\%).
Many entries in Verschuur's table have $R > 0.0017$ and these probably
represent real measurements of Zeeman splitting.

\subsection{Reliability of the HCRO \boldmath{$\HI$} Emission Results}
\label{hcroncp}

Nearly all published results in $\HI$ emission are from the HCRO
telescope.  Verschuur's bombshell was directed primarily at those
results.  Having dealt with Verschuur's criticisms, it remains to show
that our HCRO emission measurements are, in fact, correct.  Heiles (1996b)
discussed his correction procedures for the HCRO data. He also tested
these correction procedures on the North Celestial Pole, which is the
one point on the sky where, for the HCRO equatorially mounted
telescope, the telescope beam could rotate in a complete circle.

Heiles divides the data into 12 time (``Right Ascension'' or RA) bins
and measures the magnetic field strength $B_{\Vert}$ separately and
independently for each. He then Fourier analyzes the 12 results. The
Fourier terms respond differently to the beam components listed above.
Beam squint, with a two-lobed pattern on the sky, works with the first
derivative of the $\HI$ emission to produce one cycle of variation per
24 hours. Beam squash produces two cycles per 24 hours, and higher
order terms can come from the sidelobes.

These Fourier coefficients constitute {\em empirically determined}
squint and squash contributions for the NCP.  He also {\em predicted}
the squint contribution by measuring the first derivatives of the $\HI$
emission and applying the previously-measured beam squint. The two
methods gave comparable results, which shows that one can, indeed,
apply measured beam squint and squash to measured angular derivatives of
$\HI$ emission to derive -- and subtract out -- the instrumental
contribution.

Averaging over all 24 hours zeros out the contributions from beam
squint and squash, because their structure in the azimuthal direction
around beam center averages to zero. It also eliminates some, and
probably nearly all, of the sidelobe contributions. For the average
of all RAs the $V$ spectrum is an excellent fit to the derivative of
the $I$ spectrum, with $B_{\Vert} = 8.8 \pm 0.4\,\umu$G (Heiles 1996b);
this is in excellent agreement with the measurements nearby in the sky
(Heiles 1989). He also found a systematic
variation of $B_{\Vert}$ with RA from $\sim 7$ to $12\,\umu$G,
indicating the contribution of instrumental errors.  The amplitude of
the first Fourier component $\sim 2.0\,\umu$G and of the second $\sim
0.58\,\umu$G. The additional uncertainty produced by this variation,
calculated as an r.m.s., is $1.4\,\umu$G. The first Fourier component
is significantly higher than the others, while the second is comparable
to them and is probably not significant with respect to noise.

\begin{figure}[!h]
\centering
\includegraphics[width=3.5in] {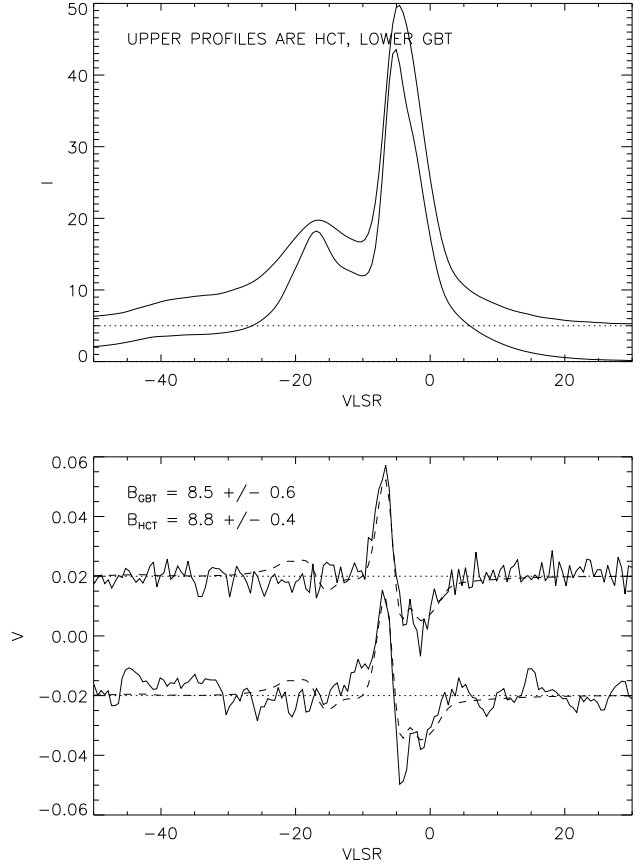}
\caption{Stokes $I$ {\bf (top)} and $V$ {\bf (bottom)} for $\HI$
emission towards the NCP for two telescopes, the HCRO and the GBT. The
upper profile in each panel is from HCRO. \label{gbt_vs_hc_rweil} }
\end{figure}

Heiles et al. (2003)
have performed a similar analysis of the North Celestial Pole using the
Green Bank Telescope. The analysis is not yet complete because the data
were taken recently. Nevertheless, the 24-hour average for the GBT is
in excellent agreement with the above HCRO results, yielding
$B_{\Vert} = 8.5 \pm 0.8\,\umu$G. Figure~\ref{gbt_vs_hc_rweil} compares
the results for the two telescopes; recall that the beam areas differ
by a factor of 16! If anything, sidelobe effects in the line wings seem
higher for the GBT spectrum.

Most of the published HCRO results did not, in fact, go through the
procedure of subtracting out the instrumental contribution.  Rather,
any position having a significant instrumental contribution, i.e.\ one
that exceeded about one third of the measured results, was not
published.  Quoted errors on the published results do not include the
instrumental contribution, so they are too small; a conservative
estimate of the instrumental error in quoted results depends on
circumstances, but is typically of order $30\%$ of the derived value.
This is relatively high, and a few quoted values may be incorrect and
even of the wrong sign.  Nevertheless, the published results should be
relatively reliable given these caveats.

All this means that HCRO reliably measured strong fields in $\HI$
emission, but not weak fields.  Thus, those measurements cannot be
used statistically, as the absorption measurements of Sect.~\ref{binabs}
can be.

\subsection{Overview of the HCRO \boldmath{$\HI$} Emission Results}
\label{magneticem}

\begin{figure}[h!]
\centering
\includegraphics[bb = 171 219 414 613,angle=90,width=11.8cm]{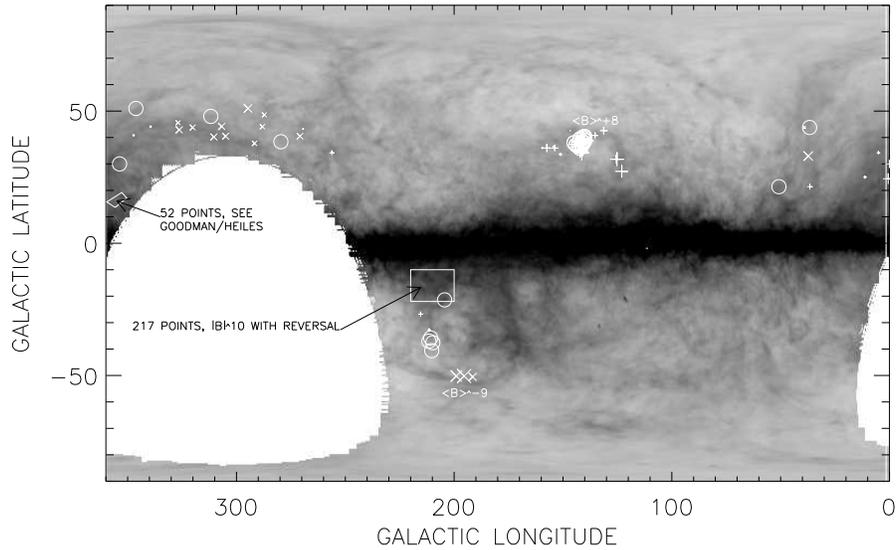}
\caption{ Detections of magnetic fields in emission from Heiles (1988,
1989), Goodman \& Heiles (1994), Myers et al. (1995) and Heiles (1997),
superposed on a map of $\HI$ in which blacker means more $\HI$. }
\label{madrid1_bw} \end{figure}

The HCRO telescope was devoted almost exclusively to Zeeman splitting
during the years before its catastrophic demise in 1993 (Heiles 1993).
It made many Zeeman splitting detections in $\HI$ emission.
Figure~\ref{madrid1_bw} shows a global map of these detections, which
are presented in five publications (Heiles 1988, 1989, Goodman \& Heiles
1994, Myers et al. 1995, Heiles 1997).
Below we present the briefest of brief summaries of each.

\begin{itemize}

\item  Heiles (1988)  mapped $\Bobs$ for 27 positions in the
vicinity of the filamentary dark cloud L204, detecting Zeeman splitting
in $\HI$ emission for all 27 and also $\HI$ self-absorption for 12
positions. This remains the best $\Bobs$-mapped example of a
well-defined, isolated dark cloud.  The $\Bobs$ exhibits correlation
with starlight polarization, CO velocities $V_\mathrm{CO}$, and the
shape of the curvy filament, implying that projection effects are
responsible for much of the structure and allowing an estimate $\Btot =
12\,\umu$G. The field dominates ram pressure from systematic flows and
also dominates the self-gravity of the molecular gas. This cloud seems
worth further study because it is well-defined with interesting
correlations, and would benefit from redoing the correlations with
better angular resolutions.

\item Heiles (1989) mapped $B_{||}$ in a number of
morphologically obvious regions, meaning high-contrast filaments. These
included several supernova or superbubble shells such as Eridanus, the
North Polar Spur, and the North Celestial Pole Loop. In every
morphologically obvious structure, the fields were strong ($| {\Bobs} |
\simgt 5\,\umu$G) and the field retained the same sign over the feature.
Magnetic pressure overwhelmingly dominates thermal pressure, and it even
dominates turbulent pressure. The paper considers the filaments to be
true filaments instead of edge-on sheets, but we wonder if this is
correct; this is an important question and needs to be resolved.  If the
structures are edge-on sheets, then the observed values $|\Bobs| \simgt
5\,\umu$G imply $\Btot \sim 10\,\umu$G from our discussion in
Sect.~\ref{bivarBN}, meaning that the field is mildly amplified in
old supernova shocks.

\item Goodman \& Heiles (1994) mapped $\Bobs$ for 52
positions in Ophiuchus, detecting it for 43 Gaussian components in 29
positions. 16 of the 43 components were in self-absorption having the
same velocity as, and therefore associated with, molecular gas.
Combining the Zeeman-splitting results with optical polarization data
allows them to determine not only $\Bobs$ but also $\Bperp$ and,
consequently, $\Btot$; it is $10.6\,\umu$G, with the field inclined to
the line of sight by $32\degr$. About half the magnetic energy is
associated with the random field component, and the magnetic and
kinetic energy densities are comparable.

\item  Myers et al. (1995) detected $\Bobs$ for 1 position in the
Draco dark cloud and 31 positions in the Ursa Major (North Celestial
Pole) loop. Magnetic and kinetic energy densities are comparable.

\hspace{5.0mm} One HCRO detection, at $(\ell, b)=(141\fdg 1, 38\fdg 8)$,
is remarkably strong, with $\Bobs = 18.9 \pm 1.8\,\umu$G. However, the
same position observed with the Effelsberg 100-m telescopes yields
the completely discrepant $\Bobs = 3.5 \pm 3.7\,\umu$G. This is a real
problem and not simply a difficulty with one of the telescopes, because
two other HCRO positions observed with Effelsberg yielded consistent
results. Given the factor 16 difference in beam area, it would seem
that there is much angular structure in $\Bobs$ at this position! But
this needs to be checked by mapping the locale with, say, the GBT.

\item Heiles (1997) mapped $\Bobs$ for 217 positions
covering $\sim 100$ deg$^2$ in the Orion/Eridanus loop region. The goal
was to develop a holistic interpretation of the magnetic field structure
on small and large size scales. The observations were interpreted  as a
large-scale ambient field distorted by the superbubble's shock, together
with smaller-scale structure produced by local perturbations. But the
match to the data is sketchy and vague, at least in part because of the
geometrical situation described in the next paragraph, so the goal was
realized only in part. Nearly all of the area mapped is permeated by a
negative field (pointing towards the observer); a small ($\sim 10$
deg$^2$) region has a uniformly positive field, which is associated with
a unique velocity component, different from those associated with the
negative field. The reversal in sign had been previously interpreted as
a toroidal field, but this may not be correct because of the different
velocity components; an alternative interpretation involves field lines
wrapped around a molecular filament by the shock front produced by the
superbubble explosions.

\hspace{5.0mm} As part of the analysis, Heiles (1997) develops a simple
geometrical model of field lines distorted by the Eridanus superbubble
shock front. For individuals who are interested in studying the magnetic
field perturbations produced by shocks, this model is worth some study
as an illustrative example of the general case. The patterns of $\Bperp$
and $\Bpar$, revealed by observations of starlight polarization and of
Zeeman splitting, are very complicated, more than one naively imagines.
They depend, firstly, on the direction of the ambient field relative to
the line of sight. They also depend on the position within the
structure. Most importantly, they also depend on which wall of the
superbubble -- the near or the far wall -- produces most of the
extinction or $\HI$ column density. The North Polar Spur, with its
easily recognizable starlight polarization effect, is a very unusual and
deceptively simple case because we see the ambient field nearly in the
plane of the sky.

\end{itemize}

\section{Importance of Magnetic Fields in Molecular Clouds}
\label{molecularclouds}

Here we will both review the observational data and focus on one of the
main reasons for observing magnetic fields in molecular clouds -- to try
to understand their role in the evolution of dense clouds and in the
star formation process. Understanding star formation is one of the
outstanding challenges of modern astrophysics. However, in spite of
significant progress in recent years, there remain unanswered
fundamental questions about the basic physics of star formation. In
particular, what drives the star formation process? The prevailing view
has been that self-gravitating clouds are supported against collapse by
magnetic fields, with ambipolar diffusion reducing support in cores and
hence driving star formation (e.g., Mouschovias and Ciolek 1999). The
other extreme is that molecular clouds are intermittent phenomena in an
interstellar medium dominated by turbulence, and the problem of cloud
support for long time periods is irrelevant (e.g., Elmegreen 2000). In
this paradigm, clouds form and disperse by the operation of compressible
turbulence (Mac Low and Klessen 2004), with clumps sometimes becoming
gravitationally bound. Turbulence then dissipates rapidly, and the cores
collapse to form stars. Hence, there are two competing models for
driving the star formation process. The fundamental issue of what drives
star formation is far from settled, on either observational or
theoretical grounds. Since the main difference between the two
star-formation scenarios listed above is the role of magnetic fields,
observations of magnetic fields in star formation regions are crucial.

Observations of magnetic fields in molecular clouds have now become a
fairly routine procedure. Great progress has been made in mapping
polarized emission from dust, and the first detections of linearly
polarized spectral lines have been made. Only the Zeeman technique has
been used for both diffuse $\HI$ and dense molecular clouds. Measuring
Zeeman splitting in molecular clouds is both easier and harder than in
the $\HI$. Instrumental effects are less important because the sources
are confined in angle so that polarized sidelobes often lie off of the
source; this makes it easier. However, molecular lines are typically
much weaker than the $\HI$ line, the frequencies are all higher, and the
Land\'e $g$ factors are somewhat smaller; although this makes it harder,
there is compensation in the form of narrower line widths and higher
field strengths in the denser molecular clouds. So progress in molecular
Zeeman measurements has been possible.

\section{Molecular Cloud Observational Results}

There has been a remarkable explosion in the observational data on
magnetic fields in molecular clouds in the last few years.  Hildebrand
and collaborators have mapped warm molecular clouds in the far infrared;
that work is reviewed by Hildebrand (2002, 2003). The JCMT SCUBA
polarimeter has been used by multiple investigators (Matthews et al.
2001; Chrysostomou et al. 2002; Wolf et al. 2003; Crutcher et al. 2004)
to map polarized dust emission at 850 $\mu$m in both warm clouds and
cool cores. The BIMA millimeter array has been used to map linearly
polarized dust and spectral line emission at 3 and 1.3 mm at
$2\arcsec-6\arcsec$  resolution (Lai et al. 2003). Crutcher (1999)
reviewed all molecular Zeeman observations made at that time and
analyzed in detail the 15 positive detections. Since then, two major
surveys of OH Zeeman have been carried out (Bourke et al. 2001; Troland
and Crutcher 2004) that have added to the total. Finally, Zeeman
measurements in OH (Fish et al. 2003; Caswell 2003, 2004) and H$_2$O
(Sarma et al. 2002) masers, which probably probe magnetic fields in
shocked molecular regions, have been made. See references to additional
results in the above papers.

Space precludes discussion of all the results. Instead, we discuss
magnetic field results for a small number of molecular clouds, chosen to
illustrate the range of the data available and the astrophysical
conclusions that may be inferred. These are a starless, low-mass core (L 183),
a region of low-mass star formation with a CO bipolar outflow (NGC 1333 IRAS4A),
a region with evidence of high-mass star formation but no $\HII$ region (DR 21 OH),
and a region with high-mass star formation and an $\HII$ region (S 106).

\begin{figure}[!h]
\centering
\includegraphics[width=2.6in,trim=0 0 0 25,clip]{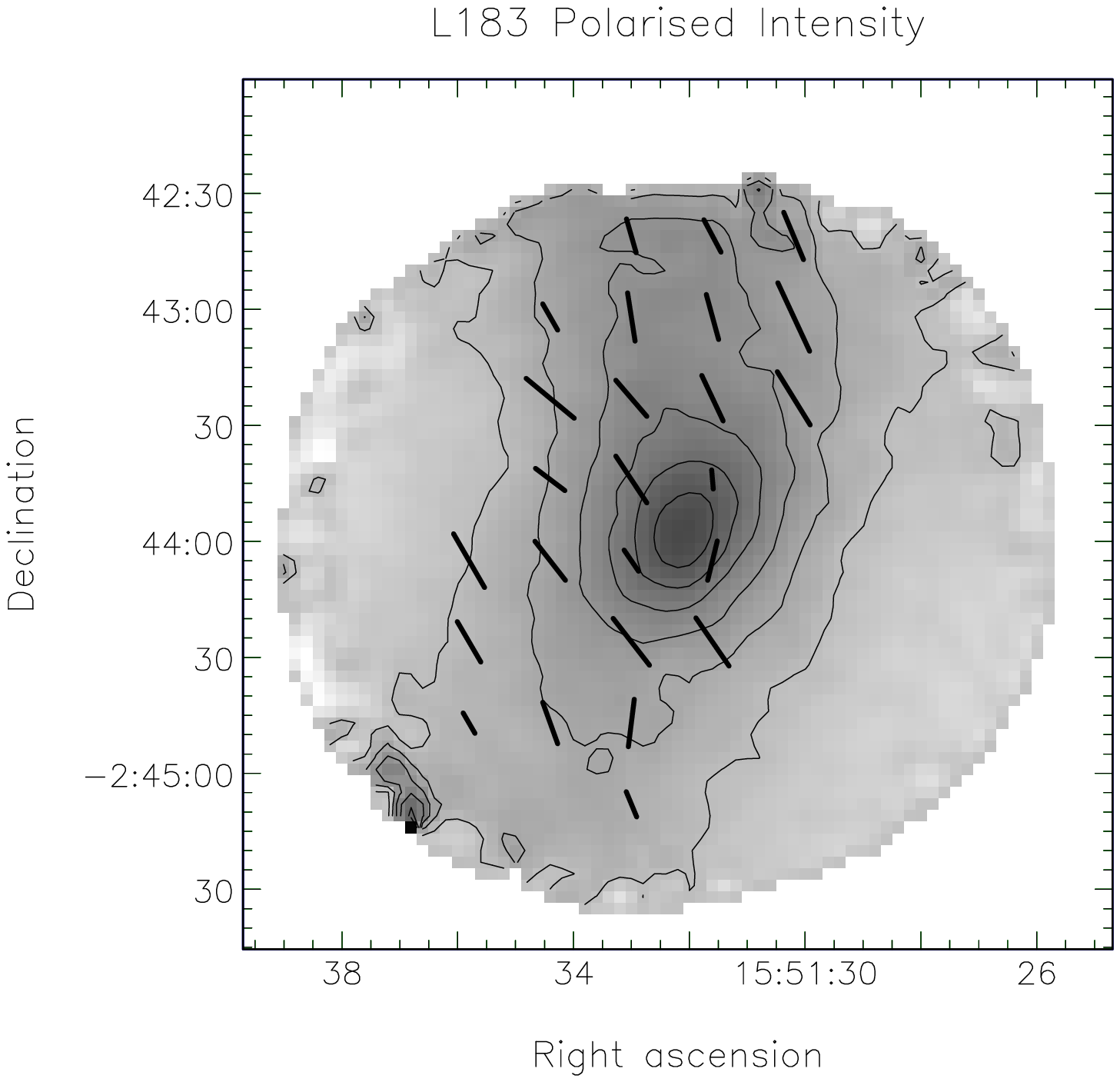}
\hspace{0in}
\includegraphics[width=1.96in]{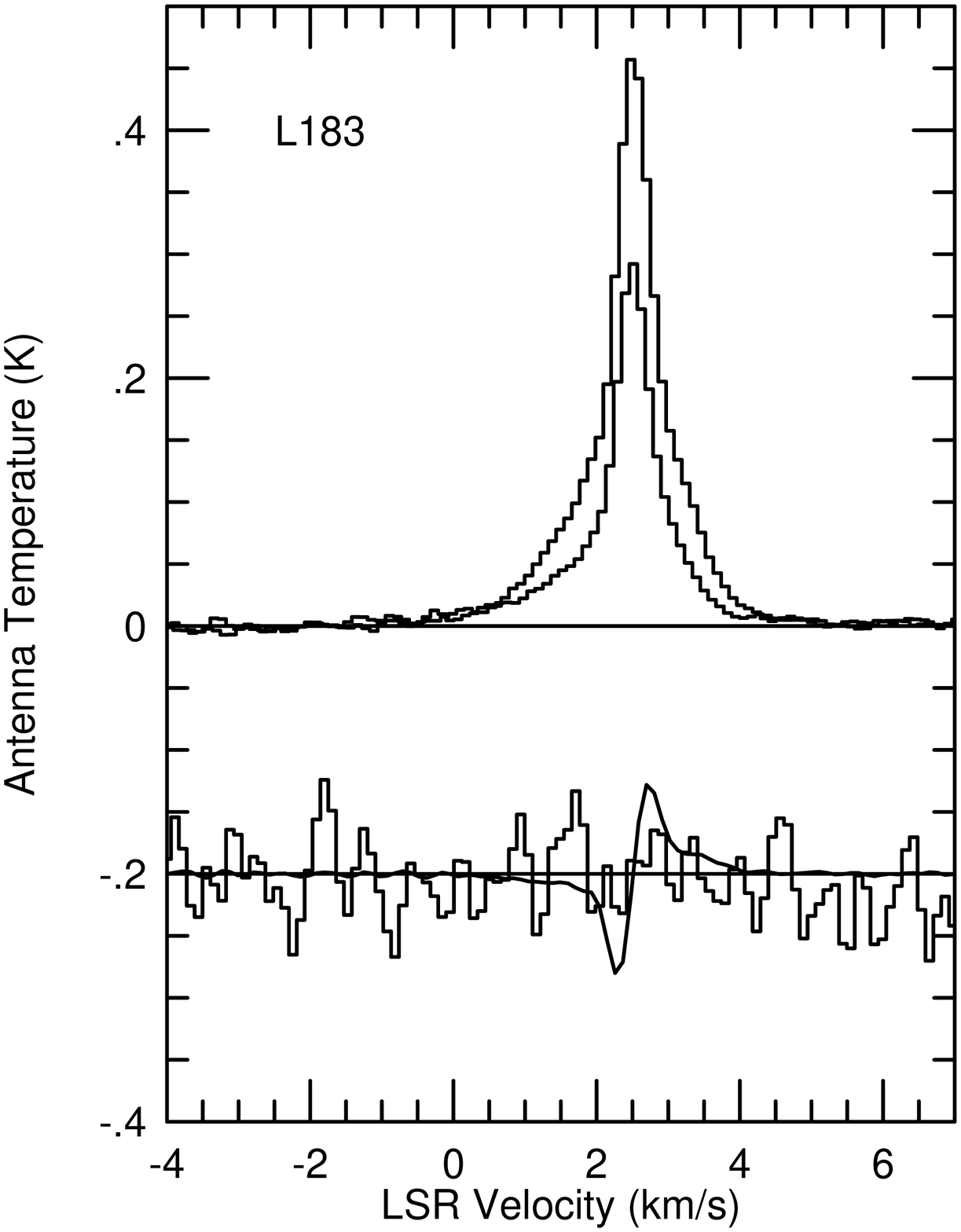}
\caption{Left: Dust polarization map of the starless core L 183. Grey-scale
and contours show the dust emission at 850 $\mu$m. Thick line segments show
the direction of the magnetic field projected on the sky; lengths are
proportional to the polarized flux. Right: OH 1665 and 1667 MHz line profiles
toward L 183. Observed data are histogram plots; the fit to Stokes $V$ in the
lower panel is a line. Top panel shows the two Stokes I spectra. Bottom panel
shows the mean Stokes V spectrum for the two lines with a $3$-$\sigma$ upper
limit fit.}
\label{L183}
\end{figure}

\subsection{The Starless Core L 183}

L 183 is a dark cloud that contains a starless core -- a dense
concentration of a few solar masses with no evidence that a protostar or
star has yet formed. Figure~\ref{L183} shows observational results for
the magnetic field; the left panel shows the SCUBA dust emission and
polarization map at 850 $\mu$m (Crutcher et al. 2004), while the right
panel shows the NRAO 43-m telescope observation of Stokes I and V
spectra of 18-cm OH lines (Crutcher et al. 1993). The dust polarization
map has an angular resolution of $21^{\prime\prime}$ and covers
$3^\prime$; the observed dust polarization position angles have been
rotated by $90^\circ$ so the line segments are in the direction of
$\Bperp$. The OH spectra were obtained with a telescope beam diameter of
$18^\prime$.

The dust polarization map samples the core of L 183, with a density of
$n(H_2) \approx 3 \times 10^5$ cm$^3$. The magnetic field is fairly
regular, in agreement with the field being strong enough to resist
turbulent twisting. But the dispersion in position angles of
$14^\circ$ is significant, implying that some
turbulent twisting is present. The angle between the projected minor
axis of the core and the mean direction of $\Bperp$ is $\sim30^\circ$.
Applying the Chandrasekhar-Fermi technique yields $\Bperp \approx 80$
$\mu$G. The OH Zeeman spectra sample a much larger area -- the extended
envelope of the L 183 core, for which $n(H_2) \approx 1 \times 10^3$
cm$^3$. The Zeeman effect is not detected to a 3-$\sigma$ upper limit of
$\Bpar < 16$ $\mu$G.

\begin{figure}
\centering
\includegraphics[width=4.3in,trim=25 0 0 0,clip] {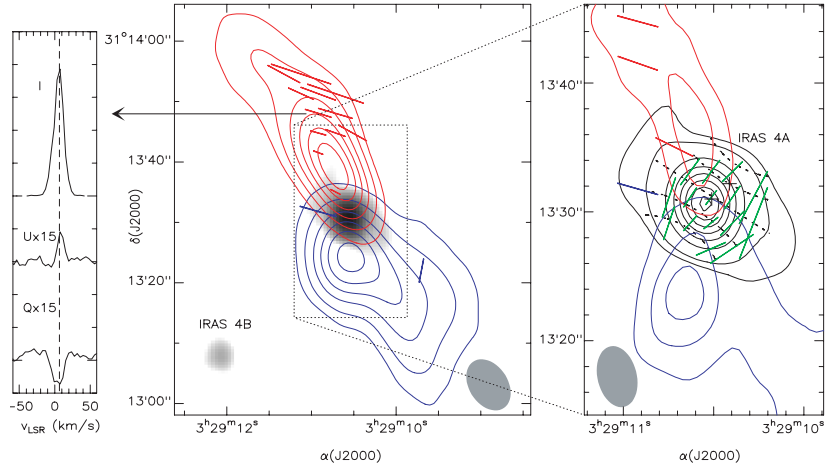}
\caption{BIMA observations of NGC 1333 IRAS4A. The middle panel shows dust
emission (greyscale) and CO 2-1 emission from the bipolar outflow (contours).
Line segments superposed on the outflow show the polarization of the line
emission. The mean Stokes I, U, and Q profiles for the northern lobe are shown
in the left panel. The right panel shows the central region dust emission
(thick contours), CO outflow (thin contours), CO polarization (black line
segments), and dust polarization (grey line segments). Dotted lines show a
possible hourglass morphology for {\bf B}.}
\label{NGC1333}
\end{figure}

\subsection{NGC 1333 IRAS4A}

NGC 1333 IRAS4A is a later stage in star formation than L 183 -- a very
young low-mass star formation region with multiple young stellar systems
and an associated molecular outflow. Figure~\ref{NGC1333} shows BIMA
observations (Girart, Crutcher, and Rao 1999) of the dust and CO outflow
emission and polarization at 1.3 mm. The line polarization is
perpendicular to the dust polarization. In the outflow, where the
direction of the velocity gradient is known, it is possible to predict
theoretically (Kylafis 1983) that the line polarization should be parallel
to $\Bperp$ and therefore perpendicular to the dust polarization, as
observed. The outflow is initially north-south, at about a $50^\circ$
angle to $\Bperp$. A successful theory of molecular outflows must
account for such a difference between {\bf B} and the outflow. However,
about $25^{\prime\prime}$ from the center the difference is only
$15^\circ$, suggesting that the field has deflected the outflow. The
morphology of the dust polarization is again smooth and suggestive of a
pinched or hourglass morphology.

 \begin{figure}
\centering
\includegraphics[width=4.5in] {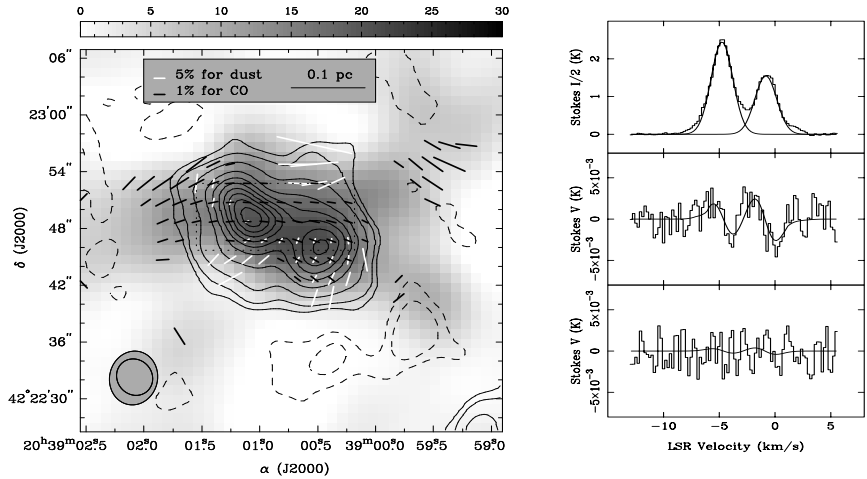}

\caption{ Left: BIMA map of the high-mass star formation region DR 21
(OH).  Contours show the 1.3-mm dust emission, grey scale shows the CO
2-1 line emission integrated over velocity, white line segments show the
dust polarization, and black line segments show the CO linear
polarization.  Right: CN 1-0 line profiles toward DR 21 (OH).  Observed
data are histogram plots, fits are lines.  Top panel shows the Stokes I
spectrum with two Gaussians fitted.  Middle panel shows the mean Stokes
V spectrum for the four hyperfine components that have strong Zeeman
splitting coefficients Z; the bottom panel shows the three components
with weak Z.  $\Bpar$ was fitted independently for the two Gaussian
lines.  The fields derived from these data are $\Bpar = -0.4 \pm 0.1$ mG
and $\Bpar = -0.7 \pm 0.1$ mG for the velocity components at $-4.7$ km
s$^{-1}$ and $-1.0$ km s$^{-1}$, respectively.}

\label{DR21OH}
\end{figure}

\subsection{DR 21 (OH)}

Figure~\ref{DR21OH} shows results for the high-mass star formation
region DR 21 (OH); the left panel shows the BIMA dust and CO emission
and polarization map at 1.3 mm (Lai, Girart, and Crutcher 2003), while
the right panel shows IRAM 30-m telescope Stokes I and V spectra of the
3-mm CN lines (Crutcher et al. 1999). In millimeter-wave dust emission
the main component of DR 21 (OH) consists of two compact cores (Woody et
al. 1989) with a total mass of $\sim$ 100 M$_\odot$. The two CN velocity
components are each centered on a different one of the two compact
cores. The region has associated masers of OH (Norris et al. 1982),
H$_2$O (Genzel and Downes 1977), and CH$_3$OH (Batrla and Menten 1988),
and high-velocity outflows powered by the two compact cores (Lai,
Girart, and Crutcher 2003). The results from the dust and CO 2-1 linear
polarization maps suggest that the magnetic field direction in DR 21
(OH) is parallel to the CO polarization and therefore parallel to the
major axis of DR 21 (OH). This could be explained by a toroidal field
produced by rotation of the double core. The strong correlation between
the CO and dust polarization suggests that magnetic fields are
remarkably uniform throughout the envelope and the cores. Both the dust
emission and the CN lines sample a density $n(H_2) \approx 1 \times
10^6$ cm$^3$. The Chandrasekhar-Fermi technique yields $\Bperp \approx
1$ mG, compared with $\Bpar = -0.4 \pm 0.1$ mG and $\Bpar = -0.7 \pm 0.1$
mG inferred from the CN Zeeman detections shown in figure~\ref{DR21OH}.
Combining these results, the total field strength $\Btot
\approx 1.1$ mG and {\bf B} is at an angle $\theta \sim 60^\circ$ to the
line of sight. However, uncertainties in $\Bperp$ and in $\Bpar$ are
sufficiently large that $\theta$ is quite uncertain.

\begin{figure}[!h]
\centering
\includegraphics[width=2.2in]{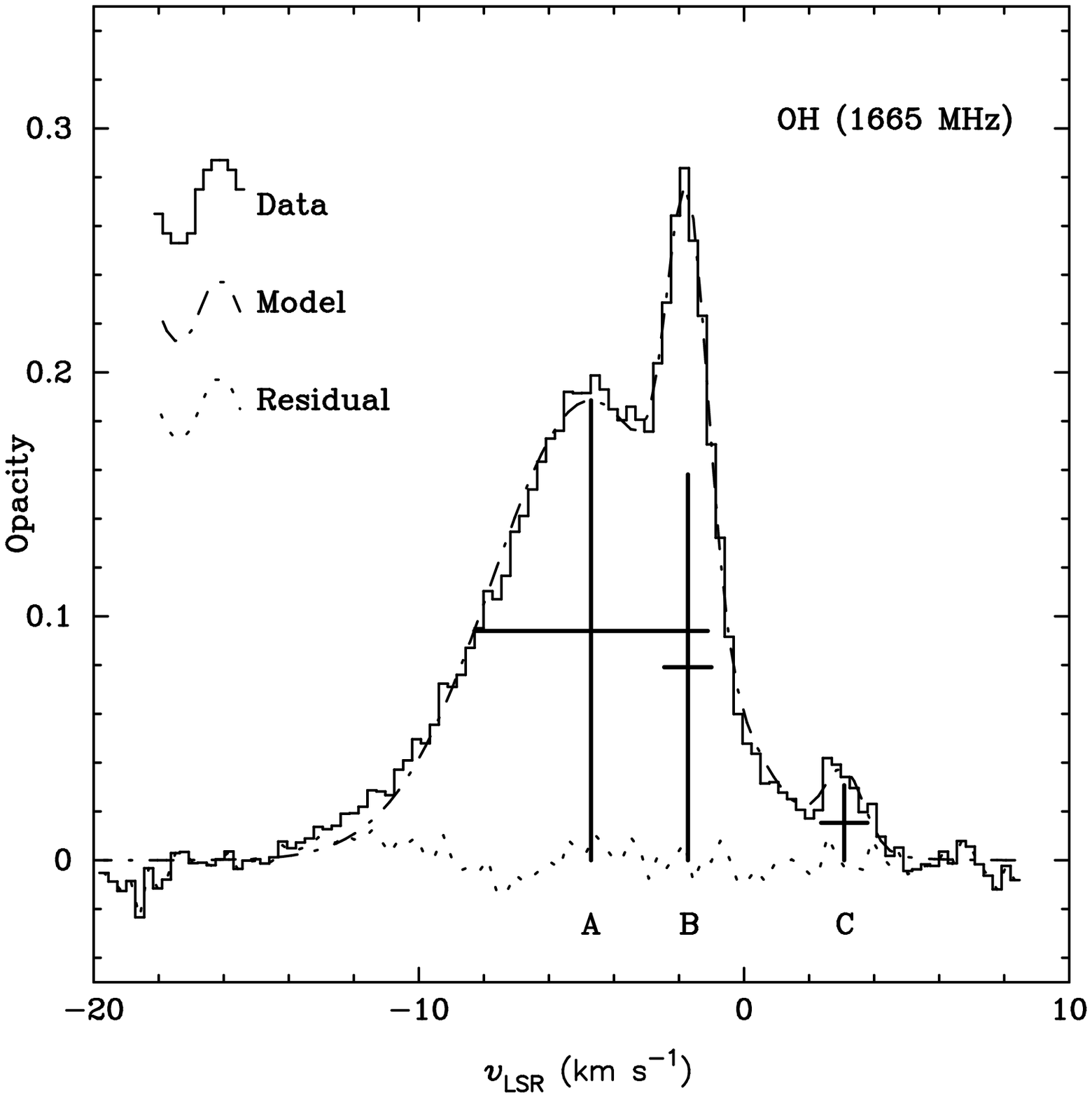}
\hspace{0in}
\includegraphics[width=2.2in]{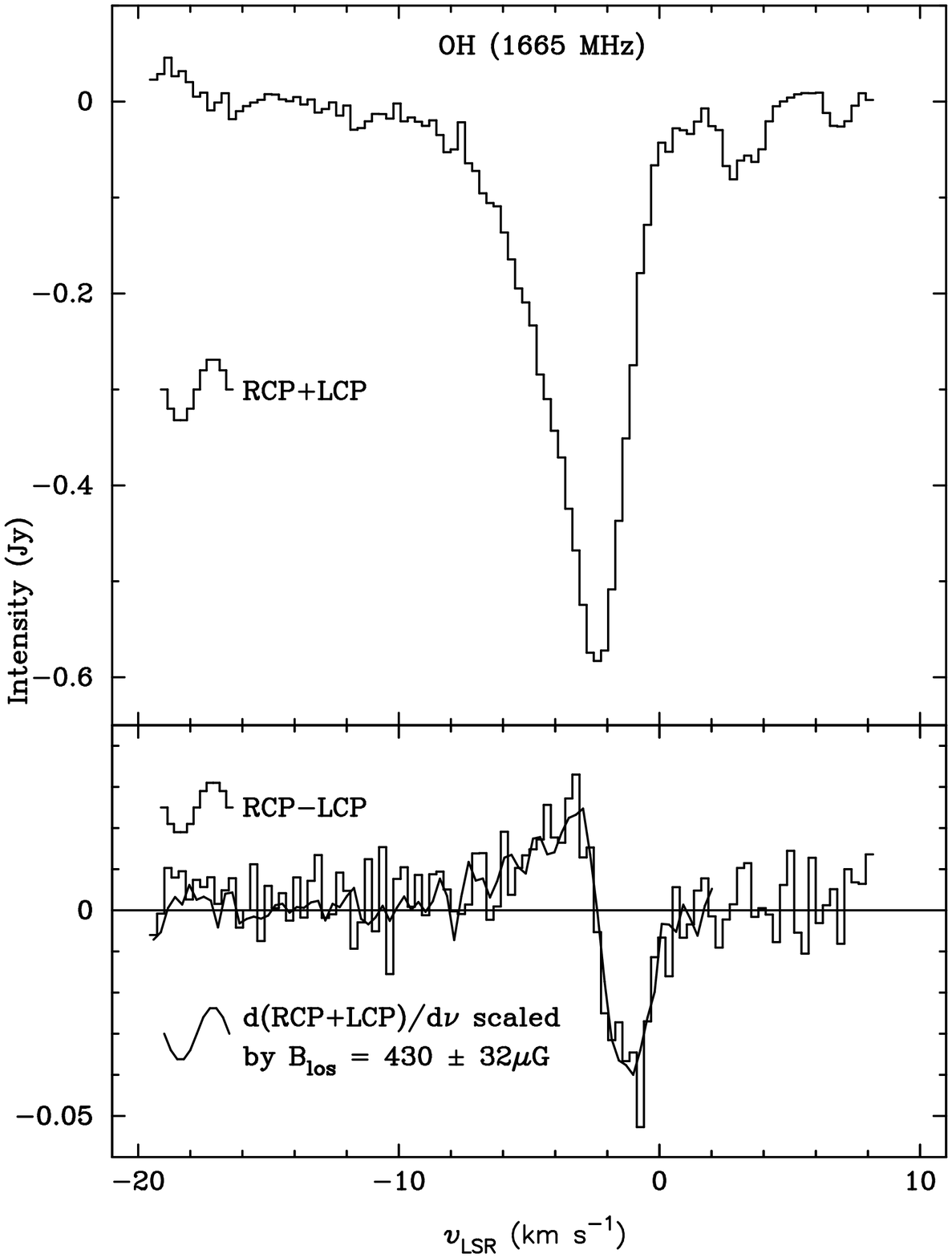}
\caption{Left: Optical depth profile for the 1665 MHz line toward S106.
Right: Stokes I and V spectra toward the position of maximum $\Bpar$ toward S106.}
\label{S106profiles}
\end{figure}

\begin{figure}[!h]
\centering
\includegraphics[width=2.2in,clip]{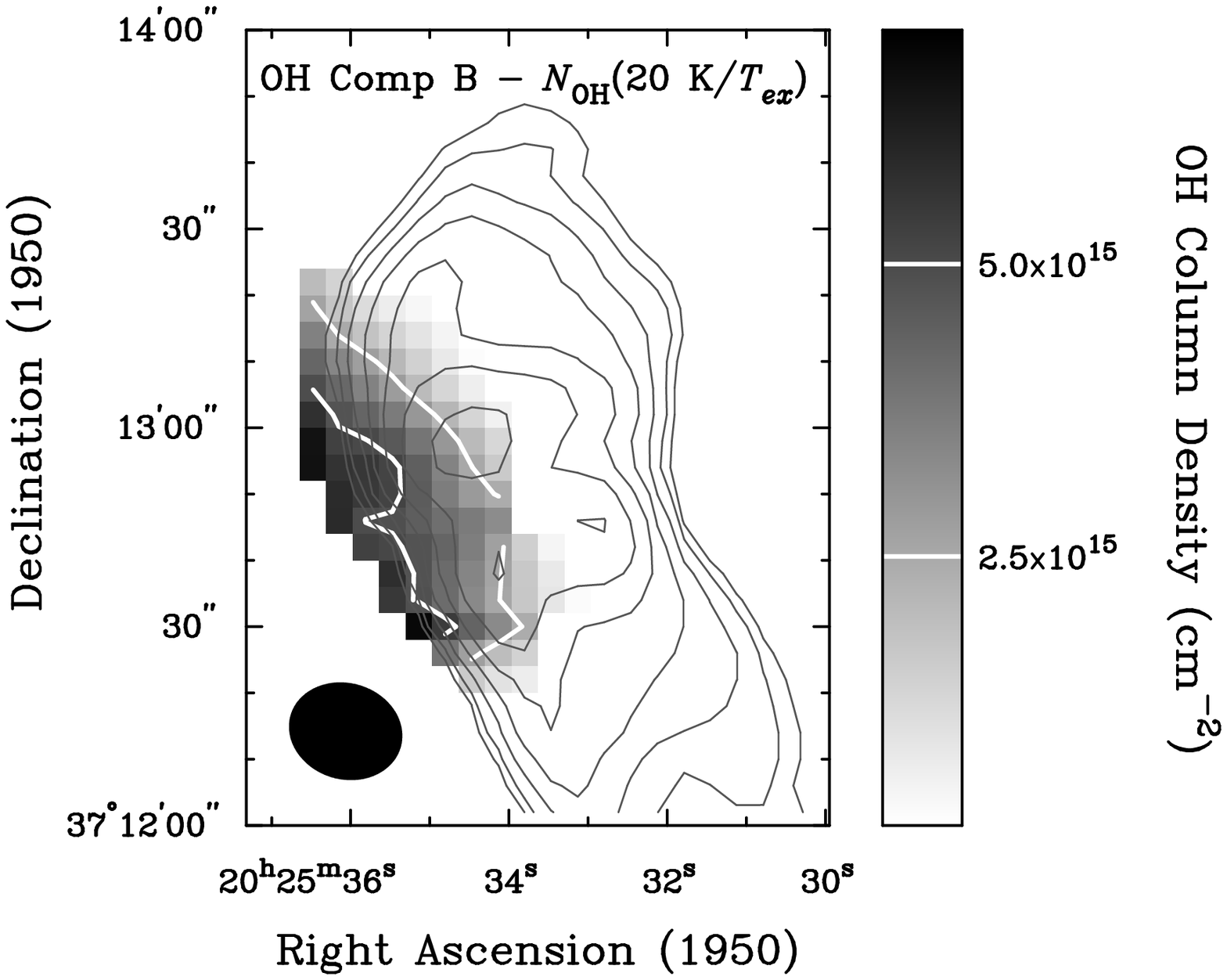}
\hspace{0in}
\includegraphics[width=2.2in,clip]{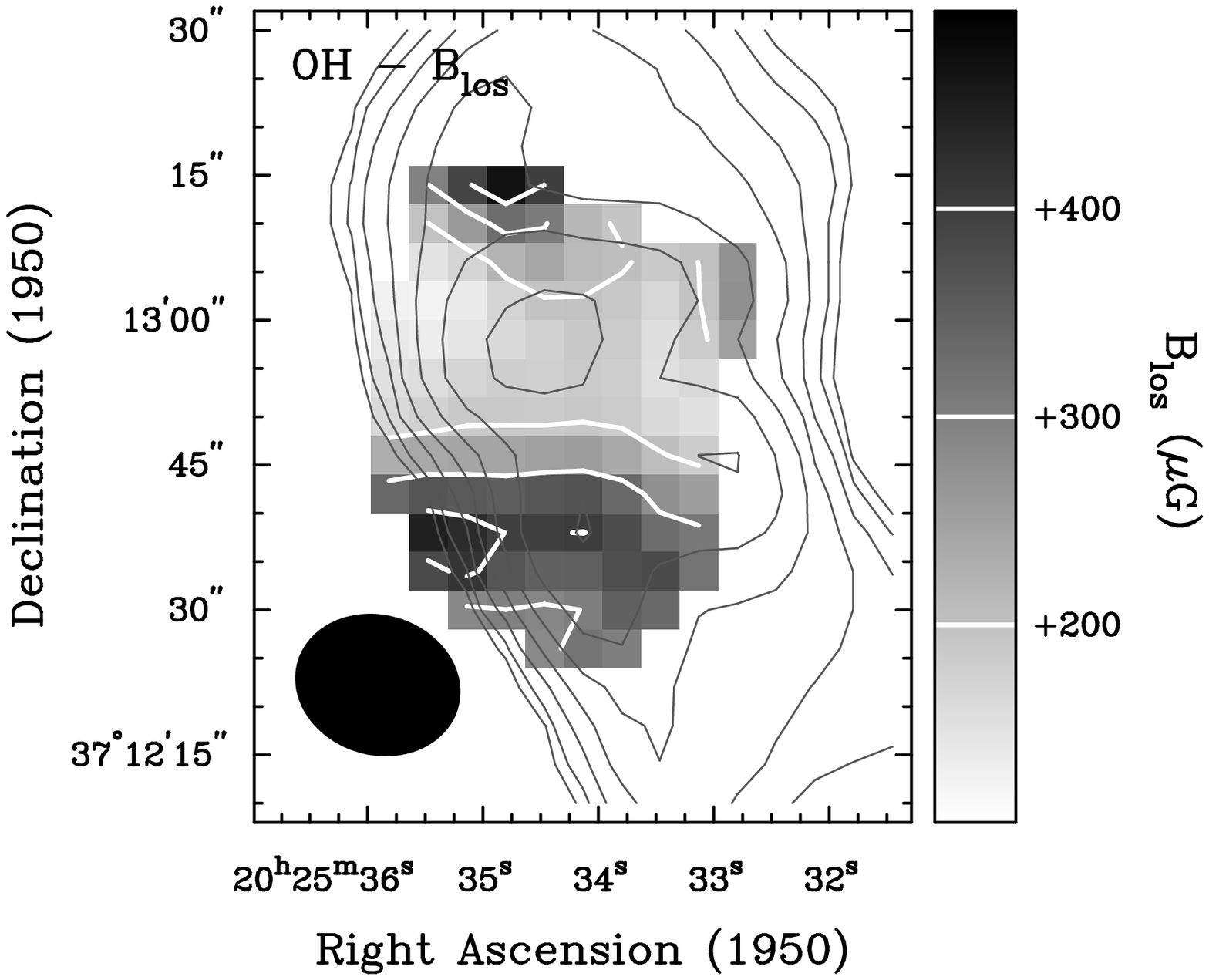}
\caption{Left: Map of N(OH) for the narrow ``B'' line component toward S106.
For $T_{ex}=50$ K (Schneider et al. 2002), contours are 1 and 2
$\times 10^{15}$ cm$^{-2}$. Right: Map of $\Bpar$ toward S106. Contours are
at 200, 300, and 400 $\mu$G.}
\label{S106maps}
\end{figure}

\subsection{S 106}
S 106 is a bipolar $\HII$ region $\sim 0.5$ pc in length embedded in an
$\sim 4$ pc diameter molecular cloud with ${\bar n(H_2)} \approx 1.4 \times 10^3$ cm$^{-3}$
and $M \approx 2000$ $M_\odot$ (Schneider et al. 2002). Roberts et al. (1995)
mapped $\Bpar$ in OH absorption lines with the VLA. Figure~\ref{S106profiles}
shows the line optical depth profile, to which three Gaussian components have
been fit. Component B is a narrow component that corresponds with the CO emission
seen over the entire molecular cloud; this is gas undisturbed by the $\HII$ region.
The broader component A arises in gas that has been shocked by the expansion of the
$\HII$ region. The Zeeman effect is seen (Fig.~\ref{S106profiles}) in component B,
so the $\Bpar$ map is of the undisturbed molecular gas and not material that has been
compressed into a shell surrounding the $\HII$ region. Figure~\ref{S106maps} shows
maps of N(OH) and $\Bpar$. The component B gas has a strong peak to the east of the
$\HII$ region, which is seen as a high-density clump in the molecular emission line
maps; Schneider et al. (2002) find $N(H_2) \approx 3 \times 10^{22}$ cm$^{22}$ for this clump.

\subsection{Maser Zeeman Observations}

OH masers are found associated with the early stage of massive star formation,
with maser spots coming from the dense ($\sim 10^7$ cm$^{-3}$) molecular envelope
surrounding the massive star. Because of their brightness, they serve as signposts
identifying sites of recently formed massive stars, and can be used to study kinematic
and physical conditions in the dense molecular material. The ground state $^2\Pi_{3/2},
J=3/2$ OH masers sometimes have clearly identifiable Zeeman pairs, that imply milligauss
magnetic field strengths. Here $\Btot$ is measured since the two Zeeman pairs are (generally)
separated. Argon et al. (2000) surveyed 91 regions with the VLA A-array in both senses of
circular polarization simultaneously, in order to identify Zeeman pairs.

Fish et al. (2003) analyzed this sample and found more than 100 Zeeman pairs in more than
50 regions. Field strengths range from $\sim 0.1$ mG to $\sim 10$ mG. They derived a
magnetic field direction for each massive star formation region and looked for correlations,
such as the correlations between maser field directions and the large-scale Galactic field
suggested by Davies (1974) based on a much smaller data set. The more complete data did not
show this correlation, which if present would have required a preservation in field direction
between the very diffuse and the very dense gas.

Excited state OH ($^2\Pi_{3/2}, J=5/2$ and $J=7/2$) maser lines were observed by
Caswell (2003, 2004). The excited-state masers tend to have fewer components and ``cleaner''
Zeeman pairs than the ground-state masers. Field strengths are similar to those found in
the ground-state maser lines.

Fiebig \& G\"{u}sten (1989) detected Zeeman splitting in the ($6_{16}-5_{23}$) H$_2$O maser
lines toward W 3, Orion KL, W49N, and S140 and inferred field strengths up to 50 mG. H$_2$O
masers probe densities $\sim 10^{8-9}$ cm$^{-3}$. Because H$_2$O does not have an unpaired
electron, the Zeeman splitting is proportional to the nuclear magneton, and only $\Bpar$
could be measured. Sarma et al. (2002) used the VLA to continue these studies, finding
$\Bpar \approx 13 - 49$ mG in four massive star formation regions. They argued that the masers
arise in C-shock regions, and that the magnetic and turbulent energies are close to equilibrium.
Sarma et al. (2001) used the VLBA to map four H$_2$O maser spots in W3 IRS5, finding that $\Bpar$
varied by a factor of three over 150 au but did not change sign. This might be expected if the
masers and magnetic field are entrained in a coherent outflow.

\section{Model Predictions and Observational Tests}

Crutcher's (1999) review of the molecular Zeeman-splitting
measurements available at that time included a detailed discussion of physical
conditions and an astrophysical discussion of the implications of the data.
He found that magnetic fields play an important role in
molecular clouds, as they do in the diffuse $\HI$ reviewed above.
Typically $\beta_\mathrm{th} \sim 0.04$ and $\beta_\mathrm{turb} \sim
1$, so the turbulent and magnetic energy densities are comparable. He also discussed
the ``mass to magnetic flux'' ratio and the scaling of $\Bobs$ with density $\rho$.
These topics will be considered in more detail below.

\subsection{Mass-to-Flux Ratio}

In contrast to the diffuse $\HI$, gravity plays an important role in molecular clouds.
From the virial theorem and assuming flux freezing, one can straightforwardly derive
the result that the ratio of gravitational to magnetic energy is independent of size.
This, in turn, means that the relative importance of gravity and magnetism is maintained.
This relative importance is measured by the ``mass to magnetic flux'' ratio ${M / \Phi}$,
which is proportional to the ratio ${ \Nperp / \Btot}$ (where $\Nperp$ is the column
density perpendicular to the sheet or disk of matter, i.e., along the magnetic field
direction for a magnetically supported cloud). We use the symbol $\mu_\mathrm{intrinsic}$
to denote ${M / \Phi}$ in units of the critical value for a slab,
$\mu_\mathrm{intrinsic} = (2 \pi G^{1/2})^{-1}$ (Nakano \& Nakamura 1978).
Then
\begin{equation}
\mu_\mathrm{intrinsic} = \ 7.6 \times 10^{-21} \frac{\Nperp(H_2)}{\Btot}.
\end{equation}

In the ambipolar diffusion model clouds are initially subcritical,
$\mu_\mathrm{intrinsic} < 1$. Ambipolar diffusion is fastest in
shielded, high-density cores, so cores become supercritical, and rapid
collapse ensues. The envelope continues to be supported by the magnetic
field. Hence, the prediction is that $\mu_\mathrm{intrinsic}$ must be $<
1$ in cloud envelopes, while in collapsing cores
$\mu_\mathrm{intrinsic}$ becomes slightly $> 1$. Hence, this model
tightly constrains $\mu_\mathrm{intrinsic}$. On the other hand, the
turbulent model imposes no direct constraints on
$\mu_\mathrm{intrinsic}$, although strong magnetic fields would resist
the formation of gravitationally bound clouds by compressible
turbulence. Also, if magnetic support is to be insufficient to prevent
collapse of self-gravitating clumps that are formed by compressible
turbulence, the field must be supercritical, $\mu_\mathrm{intrinsic} >
1$. $\mu_\mathrm{intrinsic}$ may take any value $> 1$, although of
course for turbulence models that happen to have weak magnetic fields,
clouds will be highly supercritical, $\mu_\mathrm{intrinsic} >> 1$
(Mac Low \& Klessen 2004).

If $\Btot$ is strong, clouds will have a disk morphology with \textbf{B}
along the minor axis. To properly measure $\mu_\mathrm{intrinsic}$, one
needs $B$ and $N$ along a flux tube, i.e., $\Btot$ and $\Nperp$.  We use
our discussion in Sect.~\ref{bivarBN} to relate $\mu_\mathrm{obs}$ to
$\mu_\mathrm{intrinsic}$, which is $\propto {\Nperp / \Btot}$. For a
randomly oriented assembly of sheets all having the same $\Nperp$, the
median $\Nobs$ is $2 \Nperp$. For a randomly oriented set of uniformly
strong magnetic fields, the median $\Bobs =
{\Btot / 2}$. Thus, the median value of the ratio ${\Nobs / \Bobs}$ is
$4\, {\Nperp / \Btot}$. However, it may be more appropriate to use the
mean rather than the median value:

\begin{equation}
\left\langle{\frac{M}{\Phi}}\right\rangle =
\int_{0}^{\pi/2} \frac{M_{obs} \cos \theta} {\Phi_{obs}/\cos\theta}
\sin \theta d\theta =  \int_{0}^{\pi/2} \left(\frac{M}{\Phi}\right)_{obs}
\cos^2 \theta \sin \theta d\theta =
\frac{1}{3} \left \langle \frac{M}{\Phi}\right \rangle_{obs}.
\end{equation}

Thus, the mean value of the observed ratio is three times the intrinsic
ratio, i.e.\ $\langle \Nobs / \Bobs \rangle= 3 \langle \Nperp /
\Btot \rangle$.

Crutcher (1999) listed values of $\mu_\mathrm{obs} \propto {\Nobs / 2
\Bobs}$, which are derived from observed values instead of the
intrinsic ones $\Nperp$ and $\Btot$. He included the factor of 2 for the magnetic
field, but not the additional correction factor for the column density. He noted
that such a correction would be necessary for magnetically supported clouds that
would have a disk morphology, but preferred not to apply an additional geometry
factor since the morphology of the molecular clouds was not known directly from
the observations. However, the prediction of the magnetic support model is a disk
morphology, so one must apply the column density correction to test this model.

\begin{figure}[ht!]
\centering
\includegraphics[width=4.5in,trim=10 400 200 100,clip] {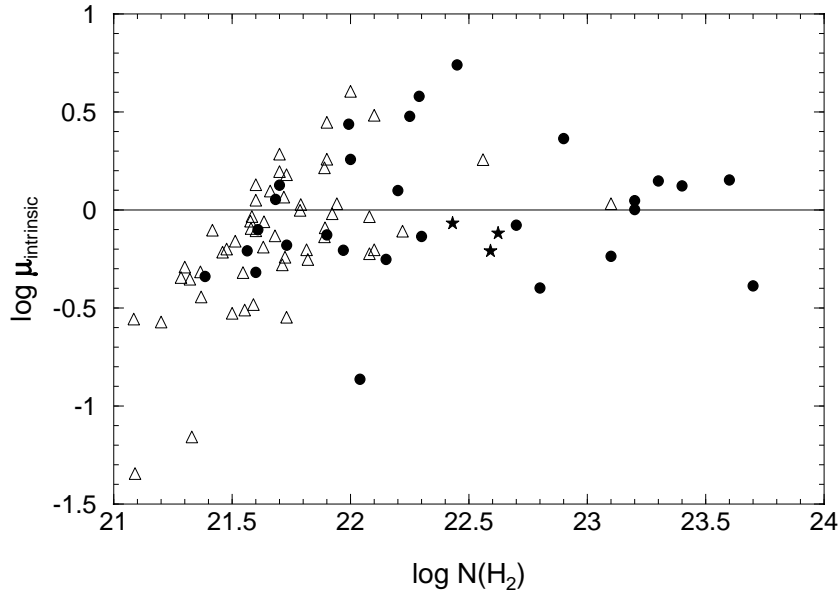}
\caption{$\mu_{intrinsic}$ are the observed mass to magnetic flux ratios,
divided by 3 to correct for projection bias, in units of the critical
values. $\mu_{intrinsic} > 1$ is supercritical, $\mu_{intrinsic} < 1$ is subcritical.
Dots are for Zeeman data with $\Bpar > 3\sigma(\Bpar)$, stars are for
Chandrasekhar-Fermi estimates of $\Bperp$, and triangles are lower limits
plotted at $\Bpar = 3\sigma(\Bpar)$. Although the statistical correction
of 1/3 for geometrical bias has been applied to each point, so that
statistically this plot should be valid, for any individual point the
true $\mu$ could be higher or lower than the plotted $\mu_{intrinsic}$. Some
of the scatter is therefore still due to geometrical projection effects. }
\label{mu}
\end{figure}

Crutcher reported the median
$\mu_\mathrm{obs,1/2}= 2.2 \pm 0.3$. We conclude that for that sample of
molecular clouds, the intrinsic and observed $\mu$ are related by
$\mu_\mathrm{intrinsic,1/2} = {\mu_\mathrm{obs,1/2} / 2}$ if we choose
the median and by $\mu_\mathrm{intrinsic} = {\mu_\mathrm{obs,1/2} /
1.5}$ for the mean.  Therefore,  $\mu_\mathrm{intrinsic,1/2}\sim 1.1$
(median) or 1.5 (mean).  This puts these clouds into the regime in which
magnetism is closely comparable to gravity. Presumably they are in
general not currently suffering gravitational collapse, because they
appear to be stable entities. (Once a core becomes supercritical, the
time scale for collapse is very short, so few cores can be at this
stage.) They are on the verge of becoming supercritical: in the absence
of external perturbations, they will gradually evolve by ambipolar
diffusion to the point where gravitational collapse can occur.
Estimates of $\mu_\mathrm{obs}$ for additional clouds may be obtained from the
OH Zeeman surveys of Bourke et al. (2001) and Troland and Crutcher (2004), and
from estimates of $\Bperp$ with the Chandrasekhar-Fermi method applied to linear
polarization maps of cores (Crutcher et al. 2004). Figure~\ref{mu} shows all of
the $\mu_{intrinsic}$ now available, where the mean value correction of 1/3 has
been used. That is, the plotted $\mu_{intrinsic} = \mu_{obs}/3$. The observations
are distributed roughly equally above and below the $\mu_{intrinsic} = 1$ line that
divides subcritical and supercritical $M/\Phi$ ratios for disk geometries. Therefore,
the data suggest that $\overline{\mu}_{intrinsic} \approx 1$; that is, the typical
mass to magnetic flux ratio is approximately critical. There is a slight indication
that for large column densities, $\overline{\mu}_{intrinsic}$ may be supercritical,
and for small column densities, subcritical.

It is also relevant to consider mass-to-flux ratios in $\HI$ clouds,
from which molecular clouds presumably form. Results from the Arecibo
Millennium Survey showed that for all of the detections, the $\mu_{obs}$
were significantly subcritical. Moreover, almost all of the
non-detections were also consistent with $\mu_{obs} < 1$. If these
points were to be plotted on Figure~\ref{mu}, they would lie to the left
of and below the $\mu_{intrinsic} = 1$ line. Hence, the $\HI$ data
suggest that the precursors to molecular clouds are subcritical, as
required by the magnetic support model.

In the ambipolar diffusion model the envelopes of dark clouds are the
regions where $M/\Phi$ remains essentially unchanged while ambipolar
diffusion drives $M/\Phi$  supercritical in the core. Hence, envelopes
of dark clouds provide a crucial test of magnetic support models --
$M/\Phi$ {\em must} be subcritical in these regions. Observations of
dark-cloud cores were carried out by  Crutcher et al. (1993), but the
$18\arcmin$ telescope beam size meant that the cores occupied a small
fraction of the beam; mainly, the envelope regions were sampled. The
result was  $\overline{\mu}_{intrinsic} \simgt 1$, rather
than the $\mu_{intrinsic} < 1$ required by magnetic support. However,
the geometrical correction to the column density was not applied; with
this correction, $\overline{\mu}_{intrinsic}$ would be slightly
subcritical, as required by the magnetic support model.

\subsection{Scaling}
\label{basu}

The scaling of $\Btot$ with density $\rho$ is usually parameterized as
$\Btot \propto \rho^\kappa$, so our discussion will be in terms of
$\kappa$. For strong magnetic fields, a cloud may be supported
perpendicular to the field, but the field provides no support along the
field. Then clouds will be disks rather than spheres. With the
assumption that self-gravity is balanced only by internal thermal
pressure along the symmetry axis $z$, $2 \pi G \rho z^2 = c^2$ (this
expression was derived for the plane-parallel or infinite thin disk case
and first applied in astrophysics by Spitzer (1942) to the structure of
the Galaxy perpendicular to the plane). Then the expression for magnetic
flux freezing ($\frac{M}{\Phi} \propto  2 \pi \rho R^2 z/\pi R^2 B$)
makes it possible to eliminate z from Spitzer's expression, yielding  $B
\propto \sqrt{\rho T}$. For an isothermal core, $\kappa = 1/2$. Detailed
calculations of the evolution of a cloud collapsing due to ambibolar
diffusion show that since the ambipolar diffusion timescale is much
shorter in a core than in an envelope, the core will become
supercritical and collapse while the envelope remains subcritical and
supported by the field. Hence, $\Btot$ in cloud envelopes remains
virtually unchanged, so at lower densities no strong correlation between
$\Btot$ and density $\rho$ is predicted, and $\kappa \sim 0$. As
ambipolar diffusion increases $M/\Phi$  in a core, $\rho$ increases
faster than $\Btot$ and $\kappa$ increases rapidly. After the core becomes
supercritical, it will collapse much more rapidly than the ambipolar
diffusion rate, and $\kappa$ continues to increase and approaches a limit
of 0.5 (Ciolek \& Basu 2000).

Once a self-gravitating clump is
formed by turbulence, if gravity exceeds both turbulent and magnetic
support, the clump will collapse rapidly, at near the free-fall rate.
Mestel \& Spitzer (1956) considered the case of a spherically
contracting cloud, for which the magnetic field was too weak to affect
the collapse morphology; they showed that $\kappa = 2/3$ for this case.
Hence, this would be the prediction for a core formed by turbulence with
no significant magnetic support against gravity. On the other hand, if virial
equilibrium is achieved between gravity and turbulence ($3GM^2/5R =
3M\sigma^2/2$), then $\rho R^2 \propto \sigma^2$. Flux freezing ($M
\propto \Phi$) gives $\rho R \propto \Btot$, so $\Btot \propto \sigma
\rho^{1/2}$ is predicted.

Determining $\kappa$ observationally can distinguish between the various scenarios.
$\kappa = 2/3$ implies a collapsing core with no significant magnetic or kinetic
support. $\kappa < 0.5$ suggests a magnetically supported cloud, with $\kappa \rightarrow
0.5$ as $M/\Phi$ goes from subcritical to supercritical. Finally, $\kappa = 1/2$ but
with an additional scaling of $\Btot$ with the turbulent velocity dispersion $\sigma$
is predicted for a core in virial equilibrium, with magnetic fields and turbulence
(or thermal motions) providing support.

At low densities $n \sim 0.1 - 100$ cm$^{-3}$, it has been clear for
some time that there is no correlation of $\Btot$ with $\rho$ (Troland
and Heiles 1986).  Crutcher's analysis of the higher density, molecular
cloud data used the observed parameters $\Nobs$ and $\Bobs$ (not the
intrinsic ones $\Nperp$ and $\Btot$).  A least squares fit showed that
$\log \Bobs \propto [\log n(H_2)]^{0.47}$, which is consistent with
ambipolar diffusion driven contraction of clouds (Fiedler \& Mouschovias
1993) or,
alternatively, with a constant Alfv\'enic Mach number $M_\mathrm{ALF}$.

\begin{figure}[h!]
\centering
\includegraphics[width=3.5in]{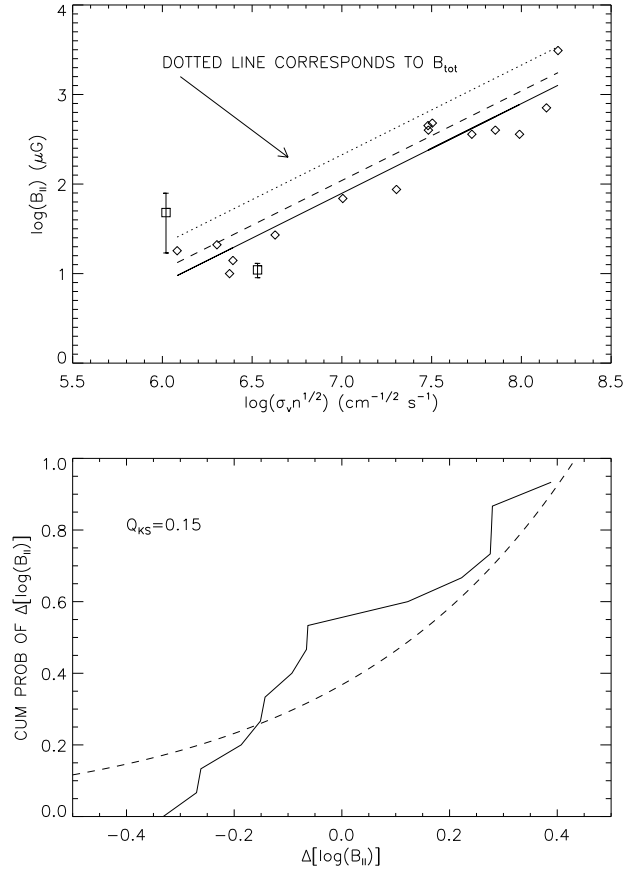}
\caption{ The {\bf top panel} shows molecular cloud data from
Crutcher (1999), together with the least-squares fit by Basu (2000)
(solid line), the correction to $\Btot$ (dotted line), and the line for
$\mu_\mathrm{intrinsic}=1$ (dashed line). The {\bf bottom panel} is the
cumulative distribution of the residuals from the fit; the dashed line
is the theoretical cumulative distribution from (\ref{cumdelta}).}
\label{basu1} \end{figure}

One year later, Basu (2000) extended Crutcher's analysis by
including the velocity dispersion in the correlation. For slablike
clouds, the combination of hydrostatic pressure equilibrium and the mass
to flux ratio yields the expected relationship from Basu's equation (3),

\begin{equation} \label{basueqn}
\Btot = ( 8 \pi)^{1/2} \sigma_v \rho^{1/2} {c_1^{1/2} \over
\mu_\mathrm{intrinsic}}
\end{equation}

\noindent where $\sigma_v$ is the velocity dispersion and $\rho$ the
mean mass density across the slab. The parameter $c_1$ relates the
midplane volume density to the mean density ($c_1 \ge 1$). Basu
replotted Crutcher's points, with the remarkable result shown in
Fig.~\ref{basu1}: the rms scatter in $\log \Bobs$ dropped by nearly a
factor of two, from Crutcher's fit with $\varDelta(\log \Bobs) \sim
0.40$, to Basu's with $\varDelta(\log \Bobs) \sim 0.23$. The data and
Basu's fit are shown in Fig.~\ref{basu1} as the diamonds and solid
line. The dashed line is the theoretical prediction from (\ref{basueqn})
for ${c_1^{1/2} / \mu_\mathrm{intrinsic}} = 1$, which is parallel
to and just little larger than the solid-line fit to the data.

The logarithmic rms dispersion $\varDelta(\log \Bobs) \sim 0.23$ is
remarkably small. This corresponds to dispersion of a factor of only 1.7
in magnetic field $\Bobs$; alternatively, because the slope is one, it
also corresponds to a factor 1.7 in $\sigma_\mathrm{v} n^{1/2}$. We
expect large variations in $\Bobs$ because of the projection factor
$\cos \theta$. We expect considerable uncertainty in the volume density
$n$, because it is estimated using a variety of rather imprecise
methods. And we also expect some cosmic scatter! The small residuals
$\varDelta(\log \Bobs) \sim 0.23$ show that this fit has physical
meaning.

Basu's result is robust with respect to the addition of new data. The
two squares with errorbars in Fig.~\ref{basu1} are new datapoints,
published after his analysis. The one with small errorbars is from OH
Zeeman splitting in L1544 (Crutcher \& Troland 2000). The one with large
errorbars is not regarded as a detection (Levin et al. 2001). Both are
consistent with Basu's fit. Although there are additional Zeeman
detections in the Bourke et al. (2001) and Troland \& Crutcher (2004)
surveys, data on $\rho$ for these clouds are not yet available; these
will provide an additional test of the robustness of the Basu result.

Basu's result convincingly shows that his model of the molecular
clouds, which is slabs in which pressure, gravity, and magnetism all
play important roles, is correct. The straightforward interpretation
from comparing the solid and dashed lines in Fig.~\ref{basu1} is that
the parameter ${c_1^{1/2} / \mu_\mathrm{obs}}$ is close to unity,
which implies both that there isn't much variation in density within
the slab and also that the mass to flux ratio is close to the critical
value.

We can go further by using the statistical discussion of
Sect.~\ref{pdfs} to relate the observed field to the total one. We
consider two results where this extension is relevant.

We now return to Basu's correlation shown in Fig.~\ref{basu1}. The
scatter of the datapoints is small, and we must ask whether it is
consistent with the statistical distribution of Sect.~\ref{univarlogb}
for $\varDelta \log \Bobs$. In particular, is the scatter too small to
be consistent with a random distribution of orientation of magnetic
field?

A least squares fit, such as done by Basu, selects the mean value of
datapoints with respect to the fitted function. The residuals of the
measured points are $\varDelta (\log \Bobs) = \log \Bobs - \langle
\log \Bobs \rangle$, where $\langle \log \Bobs \rangle$ is the mean of
the distribution. As discussed in Sect.~\ref{univarlogb}, the mean of
$\log ({\Bobs / \Btot}) = -0.43$. The distribution of the residuals
$\varDelta \log (\Bobs / \Btot)$ should follow

\begin{equation} \label{psidelta}
\psi \left( \varDelta \log {\Bobs \over \Btot} \right) = 0.85 \
 10^{\varDelta \log (\Bobs / \Btot)}
\end{equation}

\noindent We wish to compare this predicted distribution with the
observed one. Such comparisons are best done on the cumulative
distribution using the Kolmogorov-Smirnov (K-S) test. The cumulative
distribution that corresponds to (\ref{psidelta}) is

\begin{equation} \label{cumdelta}
\mbox{cum} \left( \varDelta \log {\Bobs \over \Btot} \right) = 0.368 \
 10^{\varDelta \log (\Bobs / \Btot)}
\end{equation}

The bottom panel of Fig.~\ref{basu1} shows the cumulative distribution
of the residuals as the solid curve together with the predicted one as
the dashed curve. The K-S test gives the probability $P_\mathrm{KS}$
that the two distributions are not dissimilar; here we have
$P_\mathrm{KS} = 0.15$, which although it seems small does indeed
indicate that the distributions are consistent with being identical.

We conclude that Basu's fit to Crutcher's data is statistically
consistent with a randomly oriented set of slabs. Being a least squares
fit, Basu's result provides a value $\langle \log (\Bobs / \Btot)
\rangle = -0.43$, meaning that it gives ${\Bobs / \Btot} = 0.37$.
To obtain $\Btot$ from this fit we should raise the fitted line by the
factor ${1 / 0.37} = 2.72$ (which is the base of Naperian logarithms
$e$). The dotted line in the top panel of Fig.~\ref{basu1} shows this
correction, which a factor 1.9 times higher than the dashed curve, which
represents $\mu_\mathrm{intrinsic}=1$.

In (\ref{basueqn}), this means that the factor ${c_1^{1/2} /
\mu_\mathrm{intrinsic}} = 1.9$. Above we corrected Crutcher's
observed mass-to-flux ratios to give $\mu_\mathrm{intrinsic} \sim 1.1$.
If this is accurate, then the molecular clouds are magnetically
dominated subcritical slabs with density contrast of $\sim 4$. However,
the uncertainties are such that a more appropriate summary statement is
as follows: the molecular clouds are close to the cusp of being
supercritical and have some density structure within the slab.

\subsection{Morphology}
In the magnetic support model, the dominant magnetic field means field lines
should be smooth, without irregular structure. Clouds will be thin disks or
oblate spheroids, since thermal pressure provides the only support along field
lines. The field lines should be parallel to the minor axes of clouds. Finally,
an original morphology with parallel magnetic field lines will be transformed
into an hourglass morphology since it is the tension of the bent field lines that
provides support. In the turbulent model, the magnetic field will be too weak to
resist twisting by the dominant turbulence, and field lines will not be smooth but
chaotic, with small-scale irregular structure. No correlation with cloud morphology
is expected.

Maps of dust and spectral-line linear polarization and of the Zeeman effect generally
show a regular field morphology (e.g., Figs.~\ref{L183},~\ref{NGC1333}, and~\ref{DR21OH}),
and an hourglass morphology is sometimes seen (e.g., Fig.~\ref{NGC1333}; see also
Schleuning 1998). A regular field dominating a random field and an hourglass morphology
toward cores are predictions of the strong magnetic field model. However, the magnetic
field vector projected onto the sky is not observed to be parallel to the minor axes of
starless cores as predicted by magnetic support (e.g., Fig.~\ref{L183}). Finally, even
though fairly small, the dispersion in polarization position angles is often greater than
observational errors (e.g., Fig.~\ref{L183}), implying that turbulence is producing an
irregular component to the magnetic field.

\section{Magnetic Field Observations, Present and Future} \label{future}

The field is currently in excellent health, with an unbiased
survey of absorption lines that provide statistically reliable (if
noisy) magnetic field strengths in the CNM, and a host of statistically
biased measurements with some instrumental errors in emission regions. There are
a number of molecular clouds with measured field strengths or sensitive limits,
and study of the field morphology in the plane of the sky from dust and spectral-line
linear polarization mapping is rapidly advancing. From all these measurements we
conclude that the magnetic energy density is comparable to turbulence, or larger in
some regions, and that molecular clouds are well-defined by models that incorporate
both gravity and magnetism. These results are hard-won: they require much telescope
time and, for the emission measurements, careful evaluation and correction of instrumental
contributions.

What does the future hold? In particular, what can we expect from new
instruments?

\subsection{\boldmath{$\HI$} Zeeman in Absorption}  \label{futureabs}

\subsubsection{Current Telescopes} \label{futureem}

The Arecibo Millennium survey, discussed in Sect.~\ref{binabs}, has
provided much useful statistical quantitative information about magnetic
fields in the CNM. It used nearly 1000 hours of Arecibo telescope time
to survey 79 sources in $\HI$ absorption, of which 40 (plus Cas~A from
HCRO) had useful sensitivity for Zeeman-splitting analysis. The survey
was sensitivity limited. To significantly improve the statistics, one
would want, say, four times as many sources. As we go for more sources
we inevitably go for weaker sources, so a significant improvement would
cost perhaps 10000 hours of Arecibo time. In our opinion, getting such a
time block for Zeeman splitting measurements -- indeed, for any single
scientific project -- is unlikely. And using any other telescope, with
its necessarily lower sensitivity, takes even longer. Except for special
purpose projects, we see no useful future for $\HI$ absorption Zeeman
splitting measurements using existing telescopes\footnote{This statement
applies only to diffuse $\HI$. The excellent set of Zeeman-splitting
measurements in $\HI$ associated with $\HII$ regions and supernova remnants,
made with the VLA (e.g., Brogan \& Troland 2001), can be extended to many more sources.}.

\subsubsection{The SKA} \label{skaabs}

The Square Kilometer Array (SKA) will have sensitivity about 40 times
larger than Arecibo. However, this doesn't mean that the
sensitivity-limited results go $40^2 = 1600$ times faster. The reason is
that any set of reasonable sources would all be stronger than the SKA's
system noise so integration time would be independent of source flux or
system sensitivity. In other words, 10 hours on the SKA would provide
the same limiting magnetic field strength for both a 100\,mJy source and
a much stronger 1\,Jy source. If a new
Millennium survey were performed using 1000 hours of SKA time, then
about the same number of sources could be covered as in the original
Millennium survey. This would be nice, but would probably not represent
a major scientific advance. We conclude that $\HI$ Zeeman-splitting absorption
line survey work using the SKA is unlikely to prosper.

\subsection{\boldmath{$\HI$} in Emission}

\subsubsection{Current and Future Telescopes} \label{currenttelescopes}

For $\HI$ emission, minimizing sidelobes, with their concomitant
instrumental contribution to Zeeman splitting, is paramount. This rules
out Arecibo (Heiles \& Troland 2004).
It makes two telescopes very attractive:
\begin{enumerate}

\item The Green Bank Telescope. The GBT is totally unique as a single
dish because, with its clear aperture, it should have no significant
distant sidelobes. While its sidelobes are indeed low, nevertheless we
see their effects, both in ordinary $\HI$ profiles (Stokes $I$) and
also in Zeeman splitting (Stokes $V$). We have measured these sidelobes
with complete sampling to $\sim 7\degr$ from beam center and with
incomplete sampling out to $\sim 24\degr$. This larger field shows,
surprisingly, that there seems to be little spillover from
over-illumination of the secondary. Rather, most of the Stokes $V$
effects come from within the smaller angular field. This is good news,
because it means that it might be possible to correct for their
instrumental contributions.

\hspace{5.0mm} We are currently studying the details of these
sidelobes and expect to understand them well enough to subtract out
their contribution to $\HI$ emission Stokes $V$ spectra. The degree to
which we can correct the GBT's sidelobes will determine what projects in
$\HI$ emission are feasible. Projects for which the corrections should
be easy include external galaxies other than M\,31 (because emission is
restricted in angle) and the CNM in the Milky Way (because lines are
narrow). Projects for which success should depend more seriously on
corrections include M\,31 (emission is extended, with large velocity
gradients) and the WNM in the Milky Way (lines are weak and broad). Time
will tell which projects are feasible.

\item The Allen Telescope Array. The ATA is unique among arrays in
having plenty of small baselines, which helps to provide good
brightness temperature sensitivity. At the 21-cm line the angular
resolution will about ten arcsec and the field of view some $2\fdg 5$;
a long integration on one field of view will produce a map with $10^6$
pixels. Moreover, the sidelobe properties of synthesis arrays are very
well understood, so their effects should be removable with rather good
accuracy. This will be an exciting instrument and has the potential of
revolutionizing our understanding of magnetic fields in the ISM!

\end{enumerate}

\subsection{Molecular Clouds}

\subsubsection{Current Telescopes}

The major telescopes used for Zeeman studies of molecular clouds are the VLA,
Arecibo, the IRAM 30-m, and the GBT. Including the recently completed but
unpublished survey of OH Zeeman toward dark clouds at Arecibo by Troland \& Crutcher,
there are 27 detections toward 81 positions or clouds. Because of the very large
amount of telescope time that has been expended in the OH surveys, further advances
with single-dish telescopes will probably come from Zeeman detections in CN and other
species (excited OH, SO, C$_2$S, C$_2$H, ...) that sample high-density gas rather than
from additional surveys in $\HI$ and the ground-state OH lines. The improvements to the
VLA (including especially the new correlator) that will result in the EVLA will improve
$\HI$ and OH absorption-line Zeeman mapping of clouds.

Current telescopes that have been actively used for mapping polarized dust emission
include the CSO, JCMT, and BIMA. The upgrade of the SCUBA array on the JCMT and the
combination of the BIMA and OVRO arrays into CARMA will lead to significant
improvements in sensitivity that will allow many more clouds to be mapped with higher sensitivity.
Similarly, CARMA should extend studies of linearly polarized line emission to additional
clouds. And the SMA will complement CARMA with access to higher frequencies, although with
a smaller number of antennas.

\subsubsection{Future Telescopes}

ALMA will very significantly improve the sensitivity available for dust polarization and
spectral-line linear polarization observations. With its single-dish and compact array
components, very large number of antennas, and high site, ALMA should routinely allow high
fidelity polarization mapping over extended areas of molecular clouds. For Zeeman observations
of millimeter-wave spectral lines, the improvement in sensitivity will be more modest, but
should make possible mapping of $\Bpar$ in (for example) CN in a limited number of clouds.

Although as noted above the SKA will not make it possible to significantly improve the
astrophysical results that were obtained from the Millennium Survey, its high
sensitivity will greatly increase the surface density of background continuum sources
that are strong enough for $\HI$ and OH Zeeman-splitting measurements, making
it possible to measure and map magnetic field strengths in just about any specific cloud
of interest.

\vspace{0.25in} It is a pleasure to acknowledge the pleasurable
collaborations with many Zeeman-splitting friends over the years,
especially Tom Troland.  Tim Robishaw was indispensable for the GBT
data.  Mordecai-Mark MacLow made the important suggestion regarding
equipartition of turbulence and magnetism, which we discussed in
Sect.~\ref{apdisc}.  This work was partially supported by NSF grants AST
02-05810 and AST 04-06987.

{}

\end{document}